\documentclass[sigconf]{acmart}

\newcommand{\toolname}{{\textit{DesignWeaver}}}

\renewcommand{\abstractname}{\uppercase{Abstract}}
\renewcommand{\refname}{\uppercase{References}}

\copyrightyear{2025}
\acmYear{2025}
\setcopyright{cc}
\setcctype{by-nc}
\acmConference[CHI '25]{CHI Conference on Human Factors in Computing Systems}{April 26-May 1, 2025}{Yokohama, Japan}
\acmBooktitle{CHI Conference on Human Factors in Computing Systems (CHI '25), April 26-May 1, 2025, Yokohama, Japan}
\acmDOI{10.1145/3706598.3714211}
\acmISBN{979-8-4007-1394-1/25/04}

\usepackage{listings}
\usepackage{subcaption} 
\usepackage{pgfplots}
\usepackage{xcolor}
\usepackage{array}
\usepackage{float}
\usepackage{enumitem}
\pgfplotsset{compat=1.17}

\begin{document}

\title[\toolname{}]{DesignWeaver: Dimensional Scaffolding for Text-to-Image Product Design}



\author{Sirui Tao}
\authornote{Corresponding author. Email: s1tao@ucsd.edu}
\affiliation{%
  \institution{UC San Diego}
  \city{La Jolla}
  \state{CA}
  \country{USA}
}

\author{Ivan Liang}
\affiliation{%
  \institution{UC San Diego}
  \city{La Jolla}
  \state{CA}
  \country{USA}
}

\author{Cindy Peng}
\affiliation{%
  \institution{Carnegie Mellon University}
  \city{Pittsburgh}
  \state{PA}
  \country{USA}
}

\author{Zhiqing Wang}
\affiliation{%
  \institution{UC San Diego}
  \city{La Jolla}
  \state{CA}
  \country{USA}
}

\author{Srishti Palani}
\affiliation{%
  \institution{Tableau Research}
  \city{Palo Alto}
  \state{CA}
  \country{USA}
}

\author{Steven P. Dow}
\authornote{Principle Investigator. Email: spdow@ucsd.edu}
\affiliation{%
  \institution{UC San Diego}
  \city{La Jolla}
  \state{CA}
  \country{USA}
}

\renewcommand{\shortauthors}{Sirui et al.}  


\begin{abstract}
    Generative AI has enabled novice designers to quickly create professional-looking visual representations for product concepts. However, novices have limited domain knowledge that could constrain their ability to write prompts that effectively explore a product design space. To understand how experts explore and communicate about design spaces, we conducted a formative study with 12 experienced product designers and found that experts — and their less-versed clients — often use visual references to guide co-design discussions rather than written descriptions. These insights inspired DesignWeaver, an interface that helps novices generate prompts for a text-to-image model by surfacing key product design dimensions from generated images into a palette for quick selection. In a study with 52 novices, DesignWeaver enabled participants to craft longer prompts with more domain-specific vocabularies, resulting in more diverse, innovative product designs. However, the nuanced prompts heightened participants' expectations beyond what current text-to-image models could deliver. We discuss implications for AI-based product design support tools.
\end{abstract}

\ccsdesc[500]{Human-centered computing~Empirical studies in interaction design}
\ccsdesc[500]{Human-centered computing~Text input}
\ccsdesc[300]{Human-centered computing~Graphical user interfaces}
\ccsdesc[300]{Human-centered computing~Collaborative interaction}
\ccsdesc[100]{Human-centered computing~Natural language interfaces}

\keywords{Creativity support tools, design ideation, idea management, human-AI interaction, text-to-image models}

\begin{teaserfigure}
  \includegraphics[width=\textwidth]{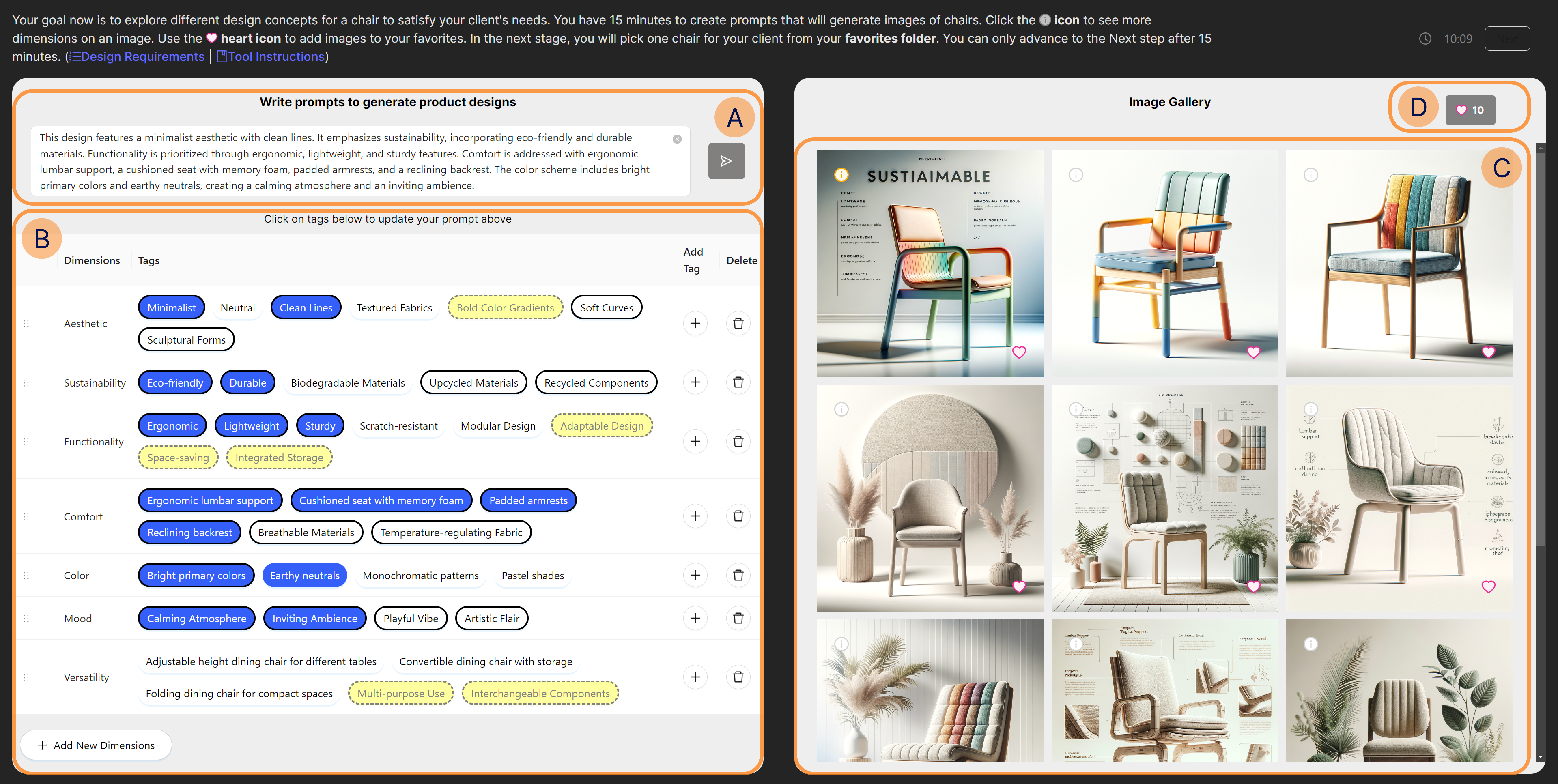}
  \caption{\toolname{}: An AI-enabled product design interface for novices. The components include (A) Prompt Box, (B) Dimension Palette, (C) Image Gallery, and (D) Favorite Folder}
  \Description{DesignWeaver: This figure shows how users will interact with the system to produce product renderings for the chair design task. An AI-enabled Product Design User Interface Designed for Novice. The components include (A) Prompt Box, (B) Dimension Palette, (C) Image Gallery, and (D) Favorite Folder}
  \label{fig:DesignWeaver_workflow}
\end{teaserfigure}

\received{September 12, 2024}        
\received[revised]{December 10, 2024}  
\received[accepted]{January 16, 2025}  

\maketitle
\section{INTRODUCTION}

Recent advancements in Generative AI (GenAI) models \cite{goodfellow2014generative, vaswani2017attention, radford2021learning, dhariwal2021diffusion} have transformed creative processes, opening new approaches to ideation and content generation \cite{davis2017quantifying, karimi2020creative}. These models show significant potential across multiple modalities, including text, images, 3D models, and video \cite{ramesh2021zero, poole2022dreamfusion, singer2022make}. In the Human-Computer Interaction (HCI) community, efforts to improve interactivity with GenAI have focused on refining prompt engineering interfaces, such as structuring prompts through familiar templates \cite{xu2024jamplate, wu2022ai, chung2023promptpaint}, studying how designers craft prompts \cite{chong2024cad, palani2024evolving}, visually rearrange AI-generated images \cite{zhang2023adding, brade2023promptify}, as well as iteratively refining Large Language Model (LLM) prompts using chaining techniques \cite{liu2022opal, di2022idea, wu2022promptchainer}. However, studies also show that supporting prompt engineering through bespoke UIs, such as through multi-modal interactions, can introduce unpredictability and hinder practical use for its increased complexity \cite{peng2024designprompt}.

Researchers have explored using text-to-image (T2I) models to generate renderings to support product design conceptualization \cite{jeon2021fashionq, liu20233dall}. T2I models can quickly produce high-fidelity visualizations that, prior to Gen AI, typically require detailed specifications about the materials and other dimensions (e.g., sizing, color, etc.). Expert designers develop domain-specific knowledge through experiences navigating diverse design scenarios, managing constraints, and mastering design principles, technical details, and available options within a given domain \cite{chong2024cad, tollestrup2023design, lawson2013design, mckenna2014adaptive}. This expertise allows designers to effectively balance high-level creative goals with technical constraints \cite{winston1970learning}. However, novice designers, even if experienced with AI prompting, often lack the domain-specific vocabulary and understanding needed to express their ideas clearly \cite{palani2021conotate, zamfirescu2023johnny}. Experts also struggle to convey tacit knowledge, and even if they express those personal and intuitive insights, GenAI may be fickle with underrepresented language in training data \cite{liu2023wants}. Experts may also benefit from structured, step-by-step prompt interactions to help wrangle unpredictable models. Current T2I models have the potential to aid both experts and novices in product design communication, but they need support to understand key terminology understood by the models. 

To understand how experts explore design spaces and communicate with their clients who generally lack domain knowledge and design expertise, we conducted a formative study and interviewed twelve experienced designers (2-20 years in professional practice) from diverse fields on how they communicate about key dimensions and options with clients. We found that designers often presented multiple alternatives to give their clients choices while maintaining creative control. Both experts and clients rely on visual representations to articulate and negotiate a vision. Based on these insights, we distill design goals for tools to help novices grapple with product design concepts, despite a lack of domain language, by emphasizing visual representations and surfacing key design dimensions and terminology.

Based on these insights, we developed \toolname{}, an interface powered by GPT-4 and DALL-E 3, designed to support novice designers through a technique we call \textit{dimensional scaffolding}. \toolname{} enables rapid iteration on text prompts by surfacing key design dimensions of user-curated images and allowing users to toggle on/off language from the "dimension palette." Key dimensions, such as geometry, style, color, and material, emerge organically through bidirectional interaction between text prompts and generated images. We hypothesize this technique helps users discover and adjust key dimensions, compensating for their lack of domain-specific knowledge, leading to more effective prompts while avoiding cognitive overload from advanced AI features.

To evaluate the effectiveness of \textit{\toolname{}}, we focused on three research questions:

\begin{enumerate} 
    \item How does dimensional scaffolding affect the \textbf{quality, length, and use of domain-specific language in text prompts} compared to a standard text-based prompting interface?
    \item How do differences in the text prompts enabled by dimensional scaffolding affect \textbf{the diversity and quality of generated product designs} compared to a standard text-based prompting interface?
    \item \textbf{How do participants engage with dimensional scaffolding} and what are their overall impressions using text-to-image models for product design?
\end{enumerate}

To investigate these research questions, we conducted a between-subjects study (n=52), where participants created a chair design based on a design brief using either \toolname{} or a baseline interface --- a standard text-based prompting interface (like ChatGPT). We collected user-generated design artifacts, system logs, survey ratings, and semi-structured interviews to gather quantitative and qualitative insights.

The study finds that \toolname{}'s \textit{dimensional scaffolding} enabled users to explore various design combinations and uncover dimensions they might not have considered with a text-based prompting interface. Participants using \toolname{} produced longer prompts with more nuance (unique domain vocabulary) from the outset. \toolname{} created a complementary interaction between text and images, leading to more nuanced prompts and better design outcomes. \toolname{} participants produced images with greater diversity as measured by CLIP-based similarity \cite{radford2021learning} and higher overall novelty as rated by blind-to-condition experts. On average, participants using \toolname{} added 3.1 new dimensions and 19.1 new tags to the palette. While this led to longer, more nuanced prompts, the \toolname{} participants issued fewer prompts overall and seemed to have greater expectations for the T2I outputs. This sometimes led to frustration (as measured by their satisfaction level with both the process and the result). We discuss how generative AI platforms might want to tamper with expectations as the underlying T2I models continue to evolve.

This paper contributes: 
\begin{itemize} 
    \item A novel system, \toolname{}, introducing dimensional scaffolding for iteratively authoring prompts and discovering key design dimensions through output inspection.
    \item An empirical study shows that \toolname{} enhances iterative prompt refinement, improves creative exploration through novel design dimensions, and fosters design innovation, with AI-based image diversity aligning closely with expert ratings on novelty.
    \item An insight that helping users author longer, more precise text prompts for GenAI may lead to frustration if not met with equally accurate image outputs.
\end{itemize}
\section{RELATED WORK}
\subsection{Challenges of Design Space Exploration}
Exploring the design space is a key part of the creative process, helping designers consider different possibilities and find the best solution, with studies highlighting its role in problem-solving and innovation \cite{maher1996modeling, dorst2001creativity}. Design space exploration involves identifying and testing various alternatives across dimensions such as form, function, aesthetics, and usability \cite{yilmaz2015design}. However, the complexity of design problems often overwhelms the exploration process, particularly for novices who may lack experience in identifying relevant dimensions and exploring trade-offs between solutions \cite{march1991exploration}. While expert designers can rely on tacit knowledge and heuristics, novices often struggle to understand the structure of the design space and how to traverse it effectively \cite{ahmed2003understanding}. Research has shown that novices focus on surface-level features and have difficulty framing problems and generating alternatives \cite{goel1992structure, atman1999comparison}. This issue is exacerbated in complex domains like product design, where multiple conflicting constraints must be balanced \cite{lawson2006how}. The formative study later aims to deepen the understanding of novices' and experts' challenges and strategies in design space exploration, building on prior research to inform the development of more effective scaffolding tools.

\subsection{Tools for Visual Design Space Exploration}
Interactive tools have been proposed to support design exploration, with some success in guiding users through multi-dimensional trade-offs \cite{frich2021digital}. Visualization tools, for instance, allow designers to navigate and explore design dimensions \cite{buxton2007sketching, kang2021metamap, suh2024luminate}. This research investigates how we might support better design exploration by novices by helping them focus on deeper, rather than surface-level, features in a design space.

Visuals play a pivotal role in design space exploration by bridging the gap between abstract ideas and tangible solutions. Prototypes, such as sketches, diagrams, and digital models, help designers articulate and explore complex design problems \cite{buxton2007sketching}. In this context, visuals act as boundary objects—shared artifacts that help designers and AI establish common ground and converge on shared interpretations to explore the design space effectively \cite{star1989structure, fischer2001articulating}. Visuals allow designers to externalize their thought processes, making tacit knowledge more explicit and facilitating better communication and collaboration \cite{gero1990design, card1999readings}. For instance, early design sketches can capture a wide range of possibilities and support iterative exploration by providing a visual reference for evaluating different design alternatives \cite{gero1990design}. Conceptual diagrams and storyboards enable designers to organize and manipulate information, which can lead to deeper insights and more informed decisions \cite{stolterman2010concept}. The iterative nature of prototyping allows designers to explore various design dimensions, including functionality, usability, and aesthetics, and discover unforeseen issues and opportunities, which can be difficult to identify through conceptual thinking alone \cite{dow2010parallel}. 

Recent advancements in Generative AI (GenAI) have significantly expanded the possibilities for generating visual content in the design process. Tools like DALL-E \cite{ramesh2021zero, ramesh2022hierarchical, betker2023improving}, Stable Diffusion \cite{rombach2022high}, and MidJourney have demonstrated tremendous potential in producing high-quality images from textual descriptions, which can help accelerate design exploration and visualization \cite{goodfellow2014generative, dhariwal2021diffusion, ramesh2022hierarchical}. These tools can generate content across various modalities, including text \cite{ouyang2022training, bai2022constitutional, touvron2023llama1, touvron2023llama2}, images \cite{ramesh2021zero, rombach2022high, saharia2022photorealistic, betker2023improving, chang2023muse, liu2023generative, liu2024logomotion}, 3D models \cite{poole2022dreamfusion, nichol2022point, gao2022get3d, lin2023magic3d, shi2023zero123++, zhou2024gala3d, liu2024one, liu2024one++}, and video \cite{singer2022make, ho2022imagen, villegas2022phenaki, kondratyuk2023videopoet, bar2024lumiere}.

While these tools have enhanced creative autonomy, they still require users to construct precise prompts, which poses a challenge for novice designers. Additionally, in the field of Human-Computer Interaction (HCI) and graphics, image-image interaction techniques such as in-painting, point-based manipulation, and sketch-based control have been explored to enhance user control over AI-generated outputs without relying solely on textual inputs \cite{iizuka2017globally, pan2023drag, zhang2023adding}. These approaches allow designers to refine outputs visually, providing more flexible and interactive methods of engaging with AI-generated images. You might generate many poor alternatives that fall into "valleys" rather than "hills," similar to simulated annealing, where the goal is to avoid suboptimal solutions by exploring a broader solution space \cite{ngoon2019dark}. Therefore, despite these advances, novice designers still encounter difficulties refining prompts or controlling the generated outputs through purely visual methods, underscoring the need for better tools that integrate visual and text-based interactions to support more intuitive design exploration. By offering a broad overview and the option to focus on details, dimensional scaffolding helps designers avoid pitfalls and discover better solutions more easily.

\subsection{Prompt Engineering for Product Design}
Prompt engineering is a critical component of GenAI interactions, as it directly influences the relevance and quality of the generated outputs. While general-purpose AI systems have made significant strides in supporting prompt crafting, they often lack the specificity required for dimension-based design exploration, where users must balance multiple design aspects like geometry, style, and functionality. Novice designers are particularly disadvantaged in these tasks, as they struggle to manage these competing design considerations. Novices struggle to construct meaningful prompts and often lack the domain-specific knowledge required to generate effective outputs \cite{zamfirescu2023johnny, palani2021active, palani2024evolving}. Novices may also find themselves overwhelmed by the complexity of generative AI systems, leading to suboptimal or irrelevant results. Even expert designers face the problem of abstraction matching and expressing tacit knowledge in text-based prompts: when the user has a well-formed intent, how do they select an utterance from the near-infinite space of naturalistic utterances that they believe the system will reliably map to a satisfactory solution? This involves “matching” the utterance to the right level of “abstraction” by specifying the utterance at a level of granularity and detail that matches the set of actions the system can take and selecting suitable words and grammar \cite{liu2023wants}. A study of how professional designers craft prompts shows that they focus more on the strategies and practices involved in prompt crafting. In contrast, novice designers often struggle with surfacing key design dimensions and lack effective prompt writing practices \cite{chong2024prompting}. This gap in prompt crafting knowledge among novices presents a significant barrier to utilizing AI tools effectively.

Efforts in HCI have focused on developing UI interfaces for prompt engineering, which assist users in crafting more effective prompts by providing structured input and output fields \cite{choi2024creativeconnect, suh2024luminate, brade2023promptify} or augmenting search \cite{son2024genquery, chen2024memovis}. For example, systems like Jamplate structure AI prompts using familiar templates, helping users generate effective outputs without fully surfacing key design dimensions \cite{xu2024jamplate}. These tools simplify prompt creation but do not provide comprehensive guidance for dimension-based exploration. In addition to structured UIs, the concept of LLM chaining has been explored, where users iteratively refine prompts and outputs by chaining multiple steps together \cite{liu2022opal, di2022idea, wu2022promptchainer}. This multi-step approach allows for more complex interactions and refinement of generated content, supporting more sophisticated design workflows. However, even these advancements do not fully address the needs of novice designers, who require more comprehensive support for balancing and exploring multiple design dimensions simultaneously. This research examines how AI can help novice designers discover the correct language for their prompts by inspecting the curated images and surfacing the key product dimensions in a palette.

\section{FORMATIVE STUDY}
This study aims to understand how designers explore the design space to meet client goals, communicate with clients to clarify preferences, and negotiate constraints to converge on a solution for further development. We conducted semi-structured, one-on-one interviews with twelve experienced designers who specialized in diverse design fields and had professional experience ranging from over two years in furniture design to more than twenty years in speculative and architectural design (see \autoref{tab:formative_participants}). 

\textbf{\textit{Participants:}} To ensure broad representation, we recruited expert designers from academia, design studios, corporate settings, and independent shops with roles in consumer electronics, medical devices, startups, bespoke furniture, and outdoor environments. Recruitment involved targeted outreach on LinkedIn, Reddit, and Discord within design-focused communities. Interested individuals signed up via an online form detailing their background and experience in creating customized client designs. We prioritized diversity in domains (e.g., furniture, product design, architecture) and experience levels (minimum of two years in the field). Our sample size aligns with formative studies in prior HCI research \cite{jeon2021fashionq, kang2021metamap, lin2024jigsaw}, offering rich insights into design generation and client communication. Continuous data analysis revealed clear convergence in participant feedback, which is shared below.

\textbf{\textit{Method:}} Semi-structured interviews lasting 30 to 60 minutes were conducted over video calls. These interviews followed a consistent protocol while allowing flexibility for participants to share unique insights. The questions focused on their design processes, client interactions, key considerations, challenges, and strategies for managing trade-offs, constraints, and client feedback. For the full list of interview questions, please refer to Appendix~\ref{appC2:fs_eip}. A qualitative analysis of the session transcripts was performed using an open coding scheme, grouping quotes into recurring themes. This approach indicated data saturation and supported the validity and robustness of our findings.

\begin{table}[htbp]
\centering
\begin{tabular}{lll}
\toprule
\textbf{Experts} & \textbf{Years of Experience} & \textbf{Design Domain(s)} \\
\midrule
E1 & 10 & Architecture, Product \\
E2 & 21 & Speculative, Architecture \\
E3 & 3 & Furniture \\
E4 & 11 & Furniture \\
E5 & 16 & Product \\
E6 & 16 & Industrial \\
E7 & 11 & Industrial \\
E8 & 3 & Product \\
E9 & 4 & Product, Furniture \\
E10 & 3 & Furniture \\
E11 & 5 & Landscape Architecture \\
E12 & 6 & Product, Industrial \\
\bottomrule
\end{tabular}
\caption{Characteristics of formative study participants.}
\label{tab:formative_participants}
\end{table}

\subsection{Insight 1: Clients Often Struggle to Articulate Specific Preferences and Rely on Visuals to Express Ideas} 
Few clients can clearly articulate their vision with specific preferences, often turning to visual aids or physical materials to communicate their ideas more effectively. As \textit{[E3]} points out, \begin{quote}\textit{“Sometimes, on top of rough sketches and photos of inspiration, clients even provide fabric swatches, color samples, or other physical materials to guide the design process.”}\end{quote} These visuals help bridge the gap between vague ideas and concrete preferences. Designers, in turn, must carefully analyze these materials to \textit{“identify key themes, categorize preferences, and highlight main priorities” (E3).} 

Designers, likewise, share visual representations with clients to allow them to get into specific details but in a colloquial manner, as \textit{[E4]} explains, \textit{“When I send a sketch to the client, they might circle parts they don’t like and say, ‘Okay, I don’t like this.’”} However, even with the help of visual aids, \textit{[E9] points out that “it can still be difficult for clients to communicate their needs fully”}. Some clients go further by bringing external inspirations, as \textit{[E8] }highlights: \textit{“Some clients bring pictures from places like France or friends, and others share videos of the furniture they want.”} Visual tools are also crucial in preventing unwanted outcomes, as \textit{[E7]} emphasizes the importance of avoiding negative surprises, \textit{“with a product that surprised users, but not in a good way,”} where clients may react with \textit{“Oh, I wasn't expecting this, and this doesn’t feel right” (E7).}

\subsection{Insight 2: Designers Present Multiple Visual Alternatives to Support Tacit Communication}
This approach typically starts with \textit{[E3] “exploring various forms and ideas before narrowing down”} to the most feasible ones. This initial exploration often leads to the presentation of \textit{[E5] “three strong options, detailing why each one is specific and discussing one standout option.”} By presenting multiple alternatives, designers can encourage more diverse and authentic feedback from clients, as this approach helps surface key design principles through comparison and exploration, ultimately fostering better design outcomes and increased confidence in the design process \cite{tohidi2006getting, dow2010parallel}. As \textit{[E6]} notes, \textit{“we impress clients by involving them throughout the brainstorming session...we display ideas on the wall and present numerous concepts that we feel strongly about.”} This method allows designers to share creative ideas and timely reference materials while quickly ruling out what works and what doesn’t for better outcomes. \textit{[E7]} further reinforces this by engaging in this activity \textit{“before moving to a full-scale model.”} This method also helps designers and clients identify key constraints, such as budget and materials, and assess where compromises can be made. \begin{quote}\textit{[E3] “By offering different design options that vary in cost, materials, and aesthetics, this helps to identify what aspects are non-negotiable for the client and where they are willing to make compromises. For instance, if a material is too expensive, we present a more affordable alternative that aligns with their design goals.”}\end{quote}

Designers often face challenges when moving into the manufacturing phase, where decisions about form and shape can cause regret. \textit{[E6]} reflects, \textit{“In hindsight, we should have designed a different product to simplify the manufacturing process…ensuring the texture and color were correct was notably difficult.”} Collaboration with engineering and manufacturing teams is crucial, as high prototyping costs, often thousands of dollars, require efficient communication and well-refined choices before implementation. The designers we interviewed consistently showed three versions with detailed breakdowns to highlight the impact of individual changes on the overall project, aiming to exceed client expectations and minimize major revisions.

\subsection{Insight 3: Designers Convey Trade-offs Across Multiple Dimensions to Educate and Manage Expectations}
During the design process, clients often bring ambitious, idealistic visions that require adjustments to align with practical realities. As noted by \textit{[E11], “[clients] will also be more idealistic. Because sometimes I find that an idea and its practicality don't go hand in hand.”} For physical products, especially furniture, space constraints often require adjustments to form and materials to ensure suitability: \textit{[E10] “I may find the need to communicate that the size constraints they’ve mentioned are not aesthetically suitable or practical for their space.”} 

Social media can further complicate these discussions, as clients often want materials they’ve seen online without considering practical factors. As highlighted by \textit{[E11], “They want the exact same materials as seen on TikTok or Instagram, and do not consider other things like their local environment or the most suitable materials.”} Budget limitations often pose a significant challenge, as clients frequently underestimate the costs of their desired designs, leading to a misalignment between their expectations and the final deliverable. \textit{[E4]} shared, “Clients often demand specific colors without realizing their budget covers less than half the cost,” highlighting how limited material availability and custom sourcing can significantly increase costs. Designers guide clients through these decisions and explain why certain choices are necessary, often convincing them to trust their expertise. As \textit{[E7]} noted, \begin{quote}\textit{“There are always situations where clients have specific preferences about the look or colors, and I need to help them understand that my choices are based on professional expertise and are the best fit for the project.”}\end{quote} 

In many cases, clients prioritize appearance over durability, making it challenging for designers to convey the importance of longevity over aesthetics for a lasting, valuable purchase. Similarly, \textit{[E10]} shared \textit{“Sometimes, clients may ask for cheaper alternatives ‘cause their initial budget is insufficient for the desired look.”} Timelines and material availability create additional obstacles, as supported by \textit{[E10]: “When I find their request impractical for their space, and when I adjust the design, it sometimes exceeds their budget and deadlines, forcing me to redo the work.”} Despite these challenges, most designers find that clients are typically satisfied with the outcome as long as the design captures their original vision.
\section{DIMENSIONAL SCAFFOLDING SYSTEM}
\subsection{Design Goals}
The Formative Study revealed a significant challenge in client-designer interactions: clients often struggle to articulate what they like or dislike about their preferred product concepts, relying on images from the web or social media rather than using technical language. This leaves designers to interpret these visuals, decipher client preferences, and explain design decisions, all while helping clients navigate trade-offs in the design process. To address this, we aim to equip designers—especially novices—with domain-specific language that enhances their ability to construct precise prompts for generating accurate reference images through large language models (LLMs). Beyond simply enabling image rendering based on less concrete or imprecise input, this approach supports a principled exploration of key design dimensions and options within the design space. By structuring these explorations and capturing critical trade-offs, designers are better positioned to systematically navigate the design space, fostering informed decision-making and thoughtful engagement with design challenges.

Based on these findings and insights from theory and practice, we aim to create a tool with the following design goals:
\begin{itemize}
    \item \textbf{\textit{Goal 1: Allow novices to absorb known preferences, requirements, and constraints.}} Establishing a context-aware foundation requires a thorough understanding of the client’s identity, vision, and specific needs, which shape and inform the design process \cite{palani2022interweave}. Gaining deep insight into the client, their value, and their project goals, requirements, and constraints allows the designer to make decisions about form, materials, or style that align with the client’s core intentions. By integrating these elements from the beginning, the designer ensures that every step of the process resonates with the client’s objectives. Collecting tangible examples and personal inspirations of what is already known is crucial to developing a clear sense of the client’s aesthetic preferences and functional requirements and effectively guiding the creative direction for the rest of the design process. Providing a framework for testing different design concepts based on these known constraints helps designers quickly explore trade-offs and refine their decisions, making it easier to compare and communicate these variations with clients.
    \item \textbf{\textit{Goal 2: Surface specific dimensions from the product design space.}} Current prompt-generation tools often lack the precision and control designers need to tailor outputs effectively, especially when clients struggle to articulate their preferences clearly. This gap highlights the need for designers to surface and refine design dimensions such as form, texture, color, and style—key elements that often emerge only through visual feedback. Sometimes, realizing too late that a different approach would have simplified the manufacturing process or prevented client dissatisfaction can be avoided by anticipating the effects of design decisions early on. By allowing designers to adjust dimensions based on visual cues or client input, the system enables flexible exploration of trade-offs, alignment with client expectations, and real-time testing of design scenarios, minimizing revisions and unforeseen issues. This flexibility is particularly valuable for novice designers, who may lack the vocabulary or experience to navigate all possible variations without such structured experimentation.
    \item \textbf{\textit{Goal 3: Enable comparison through multiple visual representations.}} Presenting multiple alternative options simultaneously allows designers to explore a broader range of possibilities and engage clients early in the design process \cite{tohidi2006getting, dow2010parallel}. By showcasing distinct visual variations that emphasize different elements, designers can demonstrate how each choice impacts the final product, facilitating comparing and contrasting options quickly and easily. Additionally, this method allows designers to justify their design choices through visual evidence, ensuring client alignment before moving into advanced stages like prototyping or full-scale modeling. Maintaining transparency throughout this process can clarify how decisions were made and ensure that both parties can easily compare and discuss the implications of each option.
    \item \textbf{\textit{Goal 4: Facilitate dimensional reasoning through trial and error.}} Novice designers often struggle to navigate complex design spaces due to a lack of domain knowledge and the language needed to communicate ideas and trade-offs. Despite access to large language models (LLMs), this gap can hinder their ability to create effective prompts or engage meaningfully with clients. An iterative feedback loop, focused on trial-and-error experimentation, provides a solution by allowing novices to refine their understanding of design-specific language and principles. Through repeated practice guided by visual references, they can explore alternatives in materials, form, and aesthetics while learning to articulate the trade-offs involved. This process builds the confidence necessary for clear client communication and fosters a deeper understanding of balancing creative vision with practical constraints. 
\end{itemize}

\subsection{DesignWeaver User Experience}

\begin{figure*}[htbp]
    \centering
    \includegraphics[width=\textwidth]{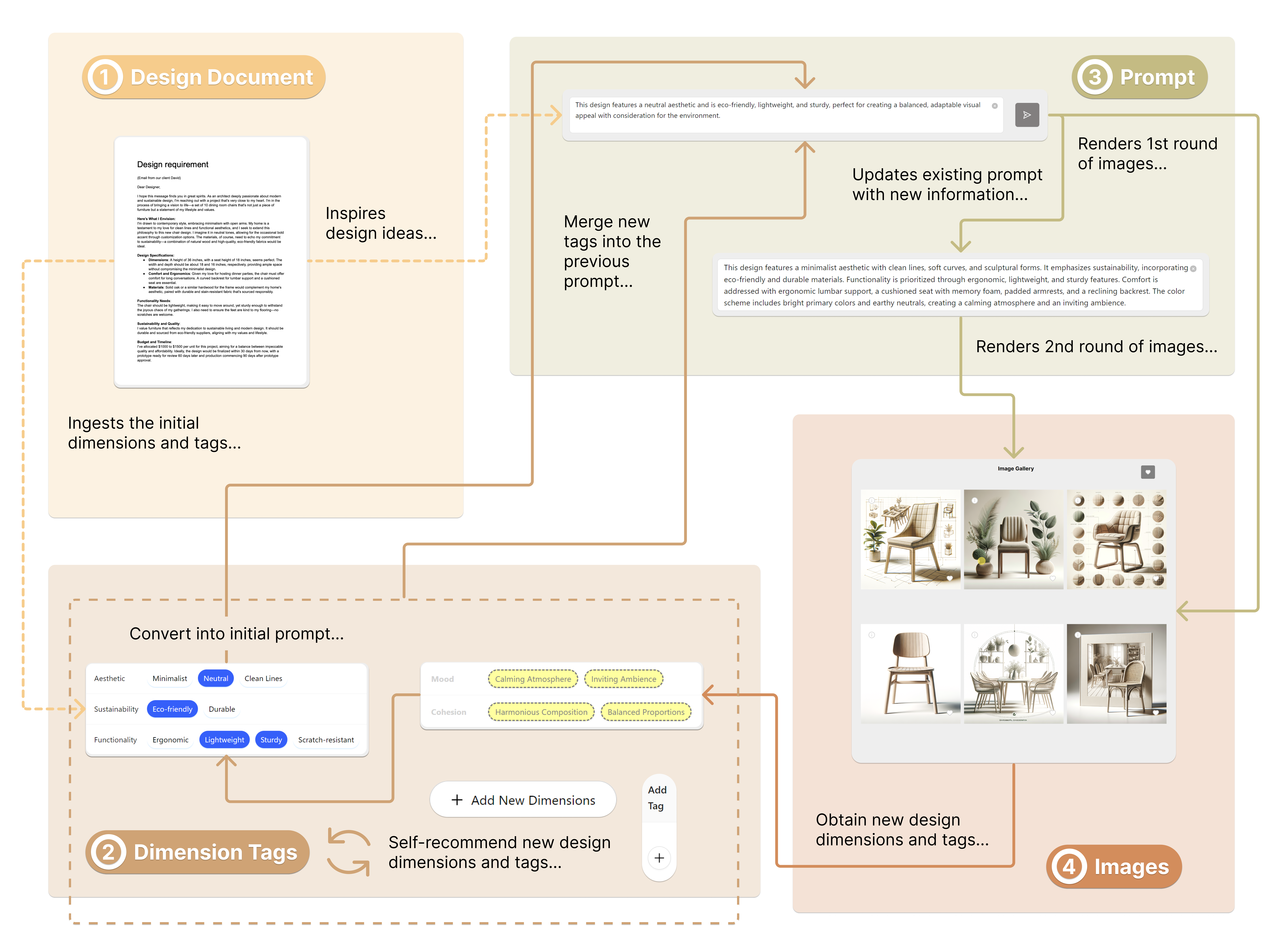}
    \caption{Overview of the iterative design process using \toolname{}. The process involves four main stages: (1) Ingest the design document to extract initial dimensions and tags, (2) Refine and recommend dimensions to generate prompts, (3) Use prompts to render and refine images, and (4) Iterate based on new dimensions and tags inspired by the generated images.}
    \label{fig:design_workflow}
    \Description{The figure shows the iterative workflow in \toolname{}, starting from the ingestion of the design document (Step 1), where initial dimensions and tags are extracted. In Step 2, dimension tags are suggested and updated based on feedback from generated images. In Step 3, prompts are generated and updated using new tags and dimensions, leading to the rendering of images (Step 4). These images further inspire new dimensions and tags, completing the cycle.}
\end{figure*}

\begin{figure*}[htbp]
    \centering
    \includegraphics[width=\textwidth]{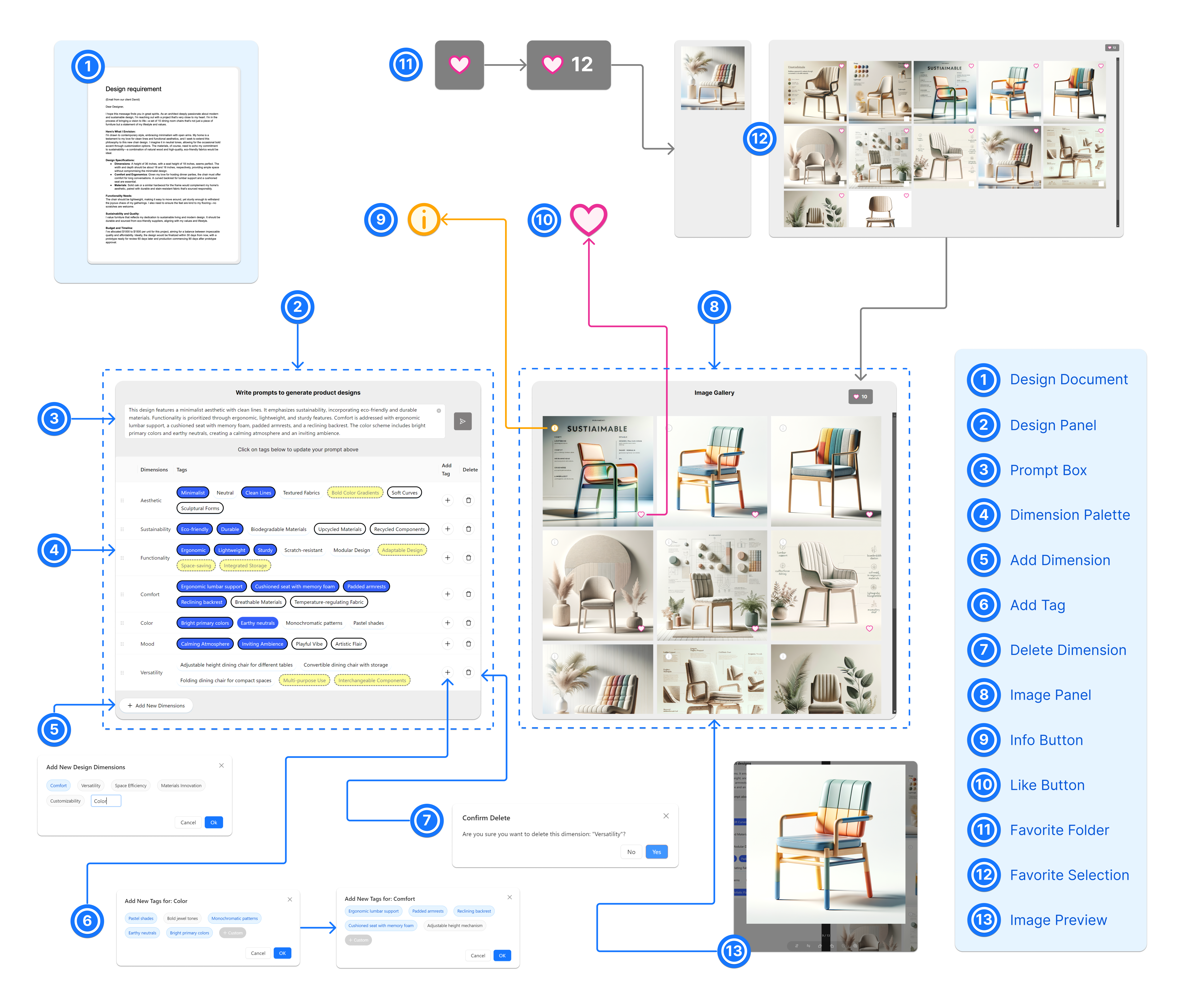}
    \caption{User Interface of DesignWeaver. The UI facilitates structured dimensional tagging and interactive exploration of AI-generated designs. Key features include a design document for guidance, a prompt box for input, a dimension palette for organizing and modifying design aspects, and an image panel displaying generated outputs. Users can add or delete dimensions, tag designs, view detailed image information, and curate favorite designs for final selection. This workflow supports iterative refinement and creativity.}
    \label{fig:designweaver-ui}
    
    \Description{
    The figure shows the user interface of the DesignWeaver tool, which facilitates structured dimensional tagging for enhanced exploration of design spaces. 

    \textbf{1. Design Document:} Displays the design requirements guiding the creative process.

    \textbf{2. Design Panel:} Provides an overview where users can write prompts to generate designs.

    \textbf{3. Prompt Box:} The area where users input prompt text for generating product designs.

    \textbf{4. Dimension Palette:} Displays the list of dimensions (e.g., aesthetic, functionality, comfort) that users can modify and explore.

    \textbf{5. Add Dimension:} Allows users to add new design dimensions to guide the generative process.

    \textbf{6. Add Tag:} Enables users to add specific tags under each dimension.

    \textbf{7. Delete Dimension:} Provides the option to remove unnecessary dimensions.

    \textbf{8. Image Panel:} Displays generated images based on the input prompts and selected dimensions.

    \textbf{9. Info Button:} Displays detailed information about each design or image.

    \textbf{10. Like Button:} Allows users to mark designs they prefer or like.

    \textbf{11. Favorite Folder:} Keeps track of liked or favorited designs.

    \textbf{12. Favorite Selection:} Shows designs or images selected as favorites for final image selection.

    \textbf{13. Image Preview:} Provides a closer look at the selected design image from the gallery.
    }
\end{figure*}

To address the aforementioned design goals, we developed \toolname{} with the following workflow shown in \autoref{fig:design_workflow}. \autoref{fig:designweaver-ui} overviews the key features.

In \toolname{}, users interact mainly through two primary panels: the Design Panel on the left and the Image Panel on the right. Before starting, an initial Design Document is uploaded. This could detail the client’s persona, vision, specifications, budget, and timeline. This document is a crucial guide throughout the design process, helping users make informed decisions while using the tool.

\subsubsection {Design Panel}
The Design Panel is divided into two sections: the Prompt Box at the top and the Dimension Palette at the bottom.
\newline
\textbf{Prompt Box.} The Prompt Box is a standard text input area that allows users to type prompts for image generation. Users could describe their design vision in their own words or use keywords as inputs for the DALL-E 3 API to function. It includes a scroll bar for longer inputs and a Send Button on the right side to submit the prompt. Prompt Box introduces two additional features. The first is an auto-complete feature, automatically completing any partially written input when activated. The second is a merging feature that combines user-written content with newly generated dimension tags (which will be explained in the next section). 
\newline
\textbf{Dimension Palette.} The Dimension Palette is the most essential feature that distinguishes \toolname{} from other tools. It helps users discover and refine design dimensions by providing categorized style tags based on keywords extracted from the uploaded Design Document. In this context, the term dimension refers to a broad design concept or aspect, such as aesthetics, sustainability, or functionality, displayed as the title of each row in the Dimension Palette. For each dimension, such as “Aesthetics,” there are associated subcategories like “modern,” “classical,” or “minimalism,” which are located within the same row. These subcategories appear as clickable style tags. Users can add new tags to any row that fits the existing dimension, delete tags, introduce new dimensions (which will create new rows), remove entire rows, and reorder them as needed. 

Users can select from a list of LLM-generated recommendations when adding new tags or creating custom tags. The exact process applies when adding new dimensions. Single clicking on a style tag within a dimension changes its color from default white to bright blue and simultaneously updates the prompt in the Prompt Box. For instance, selecting the tag “Minimalist” will automatically generate the phrase “The design embraces a minimalist aesthetic” in the Prompt Box. Users can continue to add more tags to refine the prompt further. Once satisfied, they can click the Send Button to generate images. After a brief moment, three images will appear on the right-hand side in the Image Panel, displayed in a row.

\subsubsection{Image Panel}
Users start with an empty Image Panel. Pressing the Send Button in the Prompt Box generates three images, each with a heart icon and an info icon at the bottom right. Users can click an image to zoom, flip, or expand it for a detailed preview. After generating the first set, they can modify the prompt and generate new images, repeating the process for continuous exploration.
\newline
\textbf{Like Button \& Favorite Folder.} The heart icon lets users like an image, which is then stored in the Favorite Folder at the top right for quick access. Clicking the Favorite Folder shows only liked images, helping users filter and compare them. All images remain visible in the Image Panel by default, but users can toggle between the full gallery and favorites by clicking the Favorite Folder again.
\newline
\textbf{Info Button.} Clicking the Info Icon reveals style tags on the Dimension Palette, including those used for image generation and additional ones discovered by an image understanding model (GPT-4o-mini). Tags are highlighted in yellow with dashed outlines for temporary visibility—existing tags appear bolded in yellow, while new ones have dashed outlines. A single click adds a tag and a double click activates it, turning it blue and adding it to the Prompt Box as part of the evolving prompt. The prompt updates only when a tag is activated. Users can click the Info Button again to hide all tags.
\newline
Users exploring the design space with \toolname{} switch between the Design Panel and Image Panel, forming a natural design iteration loop. The Dimension Palette displays design dimensions refined through previous prompts, helping users improve future ones. The Image Panel shows generated results, enabling visual exploration and interaction. This iterative process deepens users' understanding of the design space, revealing new design dimensions and prompting language, allowing them to craft more effective prompts and generate images that better match their design vision.

\begin{figure*}[htbp]
    \centering
    \includegraphics[width=\linewidth]{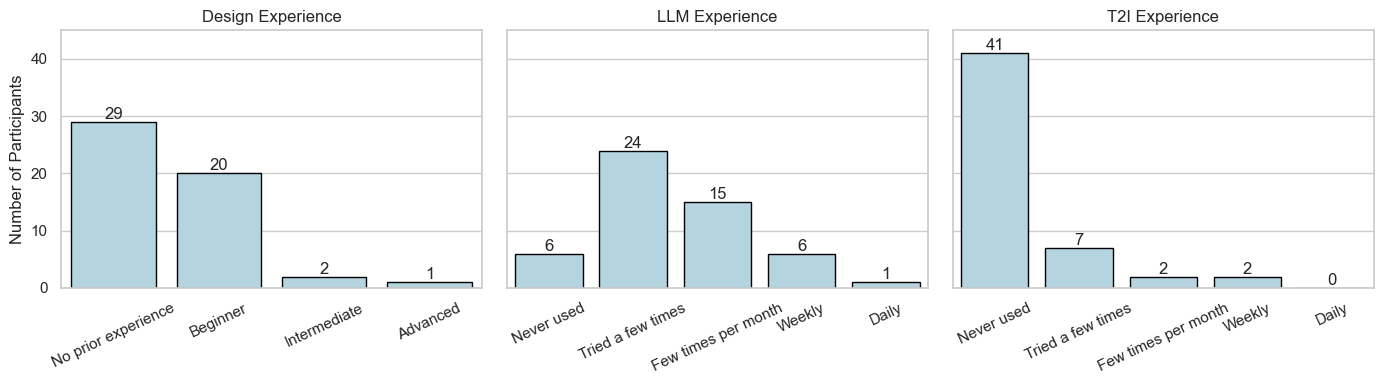}
    \caption{Number of user study participants with diverse experience in design, large language models, and text-to-image models.}
    \Description{
    The plot consists of three side-by-side bar charts showing participant experience with design, large-language models (LLM), and text-to-image (T2I) models. Each chart shares the same axes: the x-axis represents experience levels, and the y-axis shows the number of participants. 

    In the "Design Experience" chart:
    - 29 participants have no prior experience.
    - 20 participants are beginners with 0–1 year of experience.
    - 2 participants are intermediate with 2–5 years of experience.
    - 1 participant is advanced with over 5 years of experience.

    In the "LLM Experience" chart:
    - 6 participants have never used large-language models.
    - 24 participants have tried them a few times.
    - 15 participants use them a few times per month.
    - 6 participants use them weekly.
    - 1 participant uses them daily.

    In the "T2I Experience" chart:
    - 41 participants have never used text-to-image models.
    - 7 participants have tried them a few times.
    - 2 participants use them a few times per month.
    - 2 participants use them weekly.

    The bars are light blue, and numbers are displayed above or inside them for clarity. The charts include light gray horizontal grid lines for reference.
    }
    \label{fig:user_study_participant_experience}
\end{figure*}

\subsection{Implementation Details}
\toolname{} is a web application built with React and powered by a Python backend. All logged data, including prompts and tags used, are stored in Firebase, while generated images are downloaded via the Python backend and uploaded to Google FireStore. Image generation is handled using OpenAI's DALL-E 3 API. 

\subsubsection{Design Palette Initialization}
Three dimensions with corresponding design tags are extracted from the provided design document (see Appendix~\ref{appB:dd}) using GPT-4o to initialize the Dimension Palette. We instructed the model to pick the three most relevant dimensions to form the initial dimensions and design tags (see Appendix~\ref{appA1:prompts_doc_digest}). We hope the appropriate information from the design document could help users with the cold-start problem.

\subsubsection{Prompt Generation \& Update}
GPT-4o converts tags into prompt text for accurate and high-quality formatting. When tags are added or removed, GPT-4o generates and updates prompts as described in Appendix~\ref{appA2:prompts_prompt_gen_and_update}. The model compares old and new tags, ensuring tags with zero weight are removed, tags with weight one are included, and the original prompt structure is preserved as much as possible.

\subsubsection{Image Generation \& Tag Extraction}
Each time we use DALL-E 3 to generate three new images per iteration, we save their corresponding tags and prompts in Firebase (see Appendix~\ref{appA3:prompts_img_gen}). To enhance these tags, we leverage the fast processing speed of GPT-4o-mini to extract additional style tags and design dimensions from the generated images (see Appendix~\ref{appA4:prompts_tag_extraction}).

\subsubsection{New Tag \& Dimension Recommendation}
The model’s prompt incorporates existing style tags and dimensions as constraints to generate relevant recommendations. It builds on these tags to discover new dimensions by analyzing the current image, ensuring the outputs stay contextually relevant while avoiding redundant tags or dimensions (see Appendix~\ref{appA5:prompts_tag_recommendation} and \ref{appA6:prompts_dimension_extraction}).

\section{USER STUDY}
We conducted a user study to explore whether integrated dimensional scaffolding improves novice reasoning and aligns their design process with expert-level thinking. Focusing on chair design, a familiar object with rich stylistic variation, we examined how structured guidance influences participants' understanding of design dimensions, prompt construction, and final outputs. We compared our full tool with a baseline interface, offering dimensional scaffolding and structured prompts.

\subsection{Design Task}
We chose chair design for our user study because of its universal familiarity and design complexity. As an everyday object, it provides novices with a practical entry point into design processes. Chairs have long been a focus for designers and architects, especially during formative training, due to their rich history and evolution across various design movements. Their diverse design dimensions—ergonomics, aesthetics, and materials—make chairs a versatile subject that balances functional needs with creative expression \cite{fiell2005chairs}.

\subsection{Participants}

We recruited a group of 52 participants (19 to 31 years old; 36 females) in a hybrid format (5 remotely and 47 in person) from varying backgrounds and varying experience with design, large-language-model (LLM), and text-to-image-model (T2I). Regarding the respective experience level, ~\autoref{fig:user_study_participant_experience} provides a comprehensive overview.

\subsection{Baseline Condition} 

\begin{figure}[htbp]
    \centering
    \includegraphics[width=\linewidth]{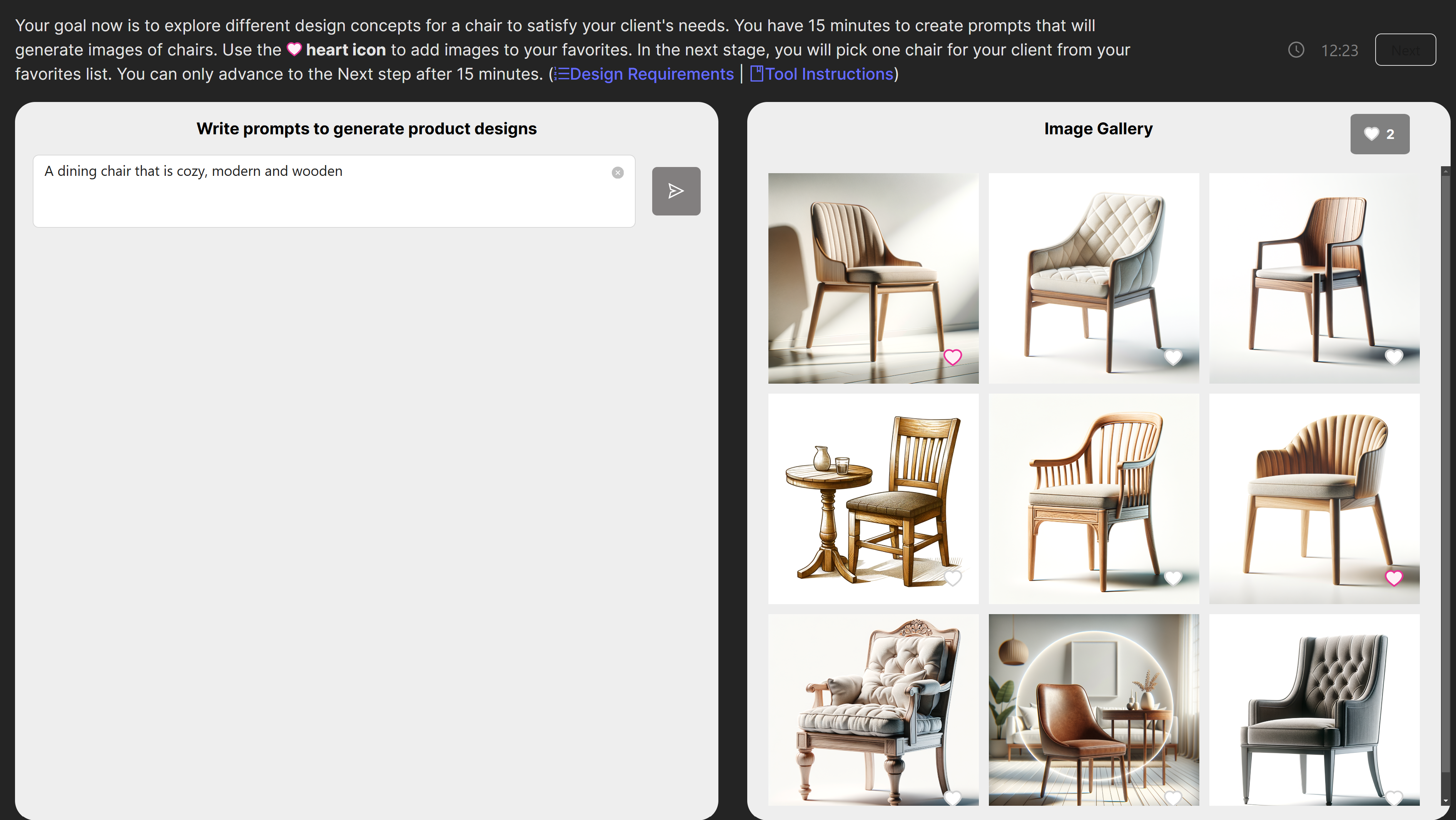} 
    \caption{The baseline interface mimics a standard text-to-image setup, excluding scaffolding components.}
    \label{fig:user_study_baseline}
    \Description{The baseline interface features a text input box labeled "A dining chair that is cozy, modern and wooden" at the top. The right panel is an image gallery with a few generated images, and 2 of them are liked. Key scaffolding features such as the Dimension Palette and info button are absent.}
\end{figure}

\begin{figure*}[htbp]
    \centering
    \includegraphics[width=\textwidth]{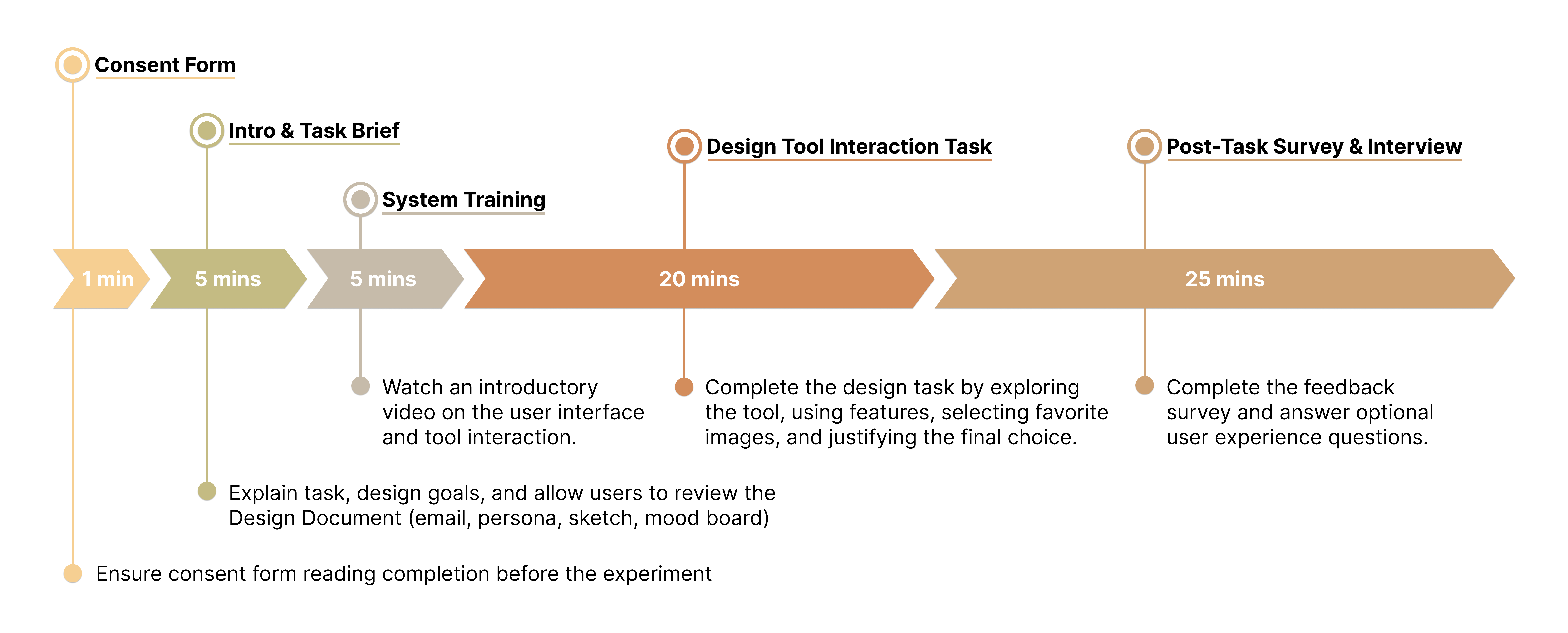}
    \caption{Workflow of the user study.}
    \label{fig:user_study_workflow}
    \Description{This figure illustrates the progression of the user study, which includes several steps such as the initial setup (1 minute), brief introduction (5 minutes), preliminary exploration (5 minutes), a main task (20 minutes), and final evaluation (25 minutes). Each arrow indicates the time allotted for that stage, helping participants understand the flow of activities.}
\end{figure*}

To assess the efficacy of dimensional scaffolding, our study compares the performance of \toolname{} with a baseline version of \toolname{}. The baseline condition removes key features such as the initial digestion of the design document, the info button on the Image Panel for extracting dimensions, and the Dimension Palette on the Design Panel, retaining only the Prompt Box, Like Button, and Favorite Folder to mimic a standard text-to-image interface (see ~\autoref{fig:user_study_baseline}). This baseline condition was chosen to isolate the specific impact of dimensional scaffolding by removing features that explicitly support design dimension extraction and organization. By mirroring a more straightforward, commonly used text-to-image interface, the baseline allows us to evaluate how these scaffolding features influence participants' ability to navigate and generate within the design space.

\subsection{Study Protocol}

Participants read the background of the user study, completed a screening survey, and reviewed the consent form. Upon consenting, participants were assigned to either the baseline group or the \toolname{} group, each receiving an introduction to their respective tools. Before beginning the tasks, a scenario overview was provided, followed by guidance on the tool features, as shown in \autoref{fig:user_study_workflow}.

\subsubsection{Consent Form (1 minute)}
Before starting this IRB-exempt experiment, participants reviewed a consent form. 

\subsubsection{Introduction to the Scenario and Familiarization with Tools (10 minutes)}
After the initial survey, participants in both groups received documents, including a client email with design requirements, a client persona with detailed information, sketches of dining chair concepts, and a mood board. Participants had 5 minutes to review these materials, which remained accessible throughout the design process. Baseline group participants received a tutorial on their baseline tool, while \toolname{} group participants were introduced to \toolname{} with additional tag selection and information reveal features. Both groups could access the tutorial slides during the design process to ensure familiarity with the tools. This introduction video stage also takes 5 minutes. 

\subsubsection{Design Exercise (20 minutes)}

Participants were then given 15 minutes, plus a 1-minute grace period, to generate images for a dining chair design. The baseline group used the tool with simple text prompts, while the \toolname{} group utilized the full tool's tag selection and information reveal features to craft prompts. Each prompt submission generated three images, and participants could view all generated images, selecting those they 'Liked' for the potential final design. An additional few minutes were provided after the exercise to finalize selections. After the time is up, they will have a few minutes to pick the final image from all the liked images.

\subsubsection{Post-Survey and Interview (20-30 minutes)}

After completing the tasks, participants filled out feedback surveys tailored to their group experience, assessing criteria for final design selection, alignment with client needs, and understanding of design dimensions. The surveys evaluated satisfaction with the tool's outputs, ease of converting ideas into prompts, and overall support in visualizing and refining concepts. They will elaborate on their Likert question response by thinking out aloud. The audio is automatically transcribed and used for later quantitative and qualitative analysis. This was followed by an in-depth interview exploring participants' experiences, design processes, tool effectiveness, and any challenges encountered. Feedback from surveys and interviews offered insights into tool interaction and highlighted areas for improvement.

\subsection{Data Collection}
We gathered data from three primary sources throughout the experiment. First, we collected the prompts used and images generated by each participant, along with the dimensions and tags used by each \toolname{} participant. We also tracked the time taken for each generation iteration. In terms of surveys, participants completed a screening survey that collected demographic and background information, including age, gender, language proficiency, and experience with design, LLMs, and T2I models. Following the experiment, participants responded to a post-experiment survey, which included Likert-scale ratings (from 1-7, Strongly Disagree to Strongly Agree) and open-ended questions covering a broad range of topics we aimed to explore.

In addition, we recorded audio transcriptions of the participants' verbal responses during the Likert scale ratings and collected elaborations through an open interview after the experiment. To evaluate the final images selected by the participants, we gathered ratings from six design experts (2-5 years of design experience), who rated novelty and alignment with the client's request on a scale from 1 to 7 (strongly dislike to strongly like).

One user from the baseline condition was excluded because a large portion of the design document was directly pasted into the prompt box, creating outliers when analyzing the prompt. This eliminated participant reduced the number of participants in the baseline group from 26 to 25.

\subsection{Data Analysis}
\subsubsection{Quantitative Data}
We compared the numerical difference between prompt length, diversity (number of design terms it contains), iteration time, and the quantitative part of the survey. Design term here is defined as a word or phrase that represents a specific aspect or dimension of a design, such as materials (e.g., "oak wood"), aesthetics (e.g., "minimalist"), functionality (e.g., "ergonomic"), or other domain-relevant attributes.

For design term extraction, we developed a customized Natural Language Processing (NLP) pipeline with four key components:

\begin{itemize}
    \item \textbf{Stopword Filtering:} Standard preprocessing (lowercasing, stopword removal, lemmatization) plus a custom list to filter non-design terms.
    \item \textbf{POS Tagging:} Extracted \textit{nouns} and \textit{adjectives} representing materials, aesthetics and key design attributes.
    \item \textbf{TF-IDF:} Ranked terms by importance, filtering out common words and highlighting key design terms.
    \item \textbf{Word Embeddings:} Used Word2Vec to ensure extracted terms aligned with a design-specific vocabulary.
\end{itemize}

We analyze the design terms from three sources: 1) the design document, 2) the prompts produced, and 3) the audio transcriptions.

Additionally, we used the ViT-B/32 CLIP model \cite{radford2021learning} to calculate both image and prompt semantic similarities for the Baseline and \toolname{} conditions across iterations. We also added Levenshtein Distances to measure the word edits in each iteration.

\subsubsection{Qualitative Data}
Two authors used Zoom's automatic transcription to transcribe and verify the recorded audio into text scripts for qualitative data, including think-aloud elaborations and open-ended user interviews. They then performed a thematic analysis of these transcripts. The insights extracted and user patterns are reported in the following result section. 

\subsection{Reliability Assessment}

We merged similar dimensions into broader meta-dimensional categories and then visualized these groupings in a bar chart and found that additional design dimensions emerged beyond the initial three (Aesthetic, Sustainability, and Functionality) that were derived from the design requirements document (see Appendix~\ref{appA1:prompts_doc_digest}). Without explicit prompt engineering, \toolname{} users organically discovered nuanced design dimensions, such as comfort, material, and durability mentioned in the design requirement, as shown in Appendix~\ref{appD:dd}. These results highlight \toolname{}'s ability to support multi-dimensional exploration and structured creativity, enabling participants to develop valid design dimensions even without professional guidance.

\section{FINDINGS}
The findings section overviews participants’ interactions with \toolname{}, highlighting how dimensional scaffolding influenced their design process and outcomes. Overall, users engaged deeply with the tool, leveraging its scaffolding features to explore design dimensions, craft detailed prompts, and iteratively refine their ideas. One notable example involved a participant curating an image with “sustainability” and “minimalist” tags, which led them to discover an overlooked product dimension—ergonomics (P13). Incorporating this dimension into their prompt generated a design that aligned with their original intent and improved functionality and client appeal. These results illustrate how \toolname{} fosters both creative exploration and the emergence of nuanced design strategies, laying the foundation for the following detailed discussion.

\subsection{Impact of Dimensional Scaffolding on Prompt Behavior}
\label{finding6.1:naunced_prompt}
\subsubsection{\toolname{} Participants Wrote Longer Prompts with Richer Design Vocabulary}

\begin{figure}[htbp]
    \centering
    \includegraphics[width=\linewidth]{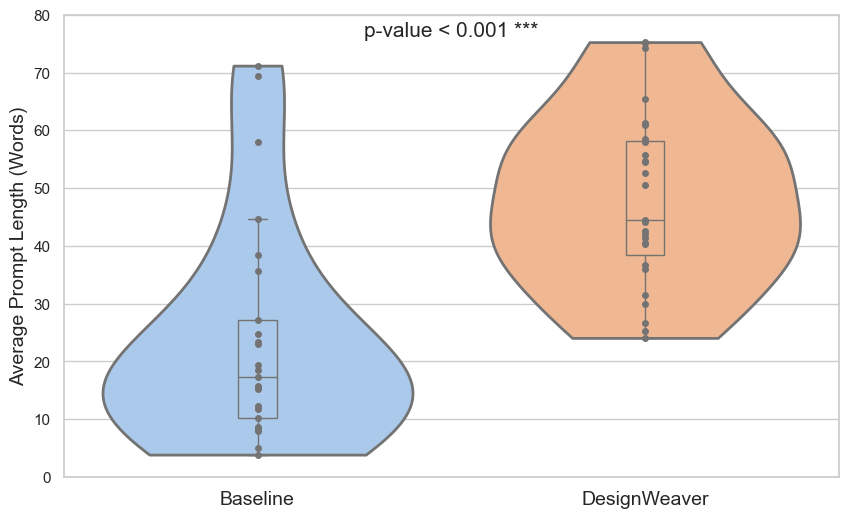}
    \caption{DesignWeaver participants issued longer prompts.}
    \Description{This plot illustrates the difference in prompt length between the two groups, highlighting the effectiveness of \toolname{} in encouraging more detailed prompts.}
    \label{fig:findings_prompt_length}
\end{figure}

\begin{figure}[htbp]
    \centering
    \includegraphics[width=\linewidth]{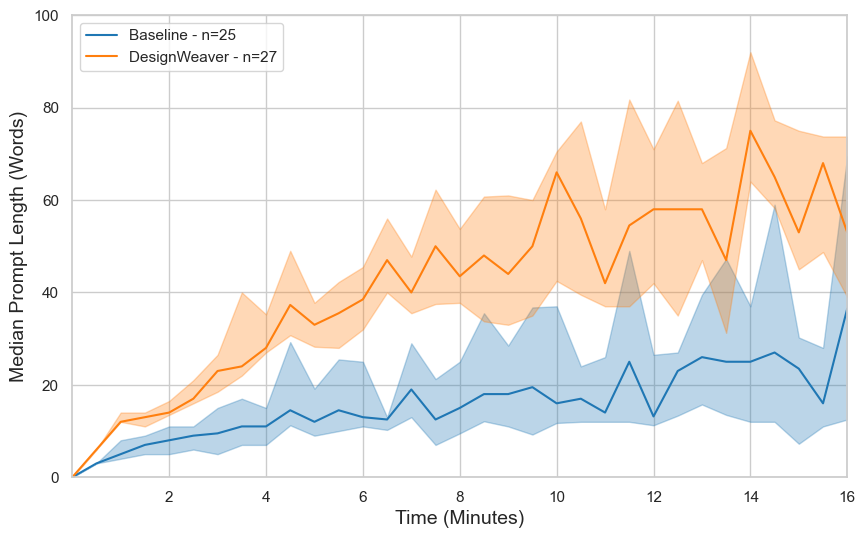}
    \caption{DesignWeaver participants consistently had higher median prompt lengths over time.}
    \Description{This graph tracks how prompt length evolved for both groups during the session.}
    \label{fig:findings_prompt_length_over_time}
\end{figure}

Prompts generated in the \toolname{} condition were more developed compared to those from the Baseline condition. These prompts reflected a deeper consideration of design dimensions such as aesthetics, functionality, and sustainability. The Baseline prompt described a dining chair with basic features like material, color, and comfort. In contrast, the \toolname{} condition prompted a more nuanced design, incorporating contemporary aesthetics, eco-friendly materials, ergonomic considerations, and playful geometry, showcasing a richer understanding of design dimensions (as shown in Appendix~\ref{appD:pc}).

Participants using \toolname{} produced significantly longer prompts (M = 48.22, SD = 15.03) compared to the Baseline group (M = 23.73, SD = 19.12), with a statistically significant difference ($U = 97.0, p < 0.001$), as illustrated in \autoref{fig:findings_prompt_length}. Analysis of prompt length over time showed that participants in the DesignWeaver group consistently expanded their prompts, with a steady increase in median prompt length throughout the session (see \autoref{fig:findings_prompt_length_over_time}). In contrast, participants in the Baseline group reached a point of stabilization earlier, suggesting a more constrained approach to prompt development and limited iterative exploration. 

\begin{figure}[htbp]
    \centering
    \includegraphics[width=\linewidth]{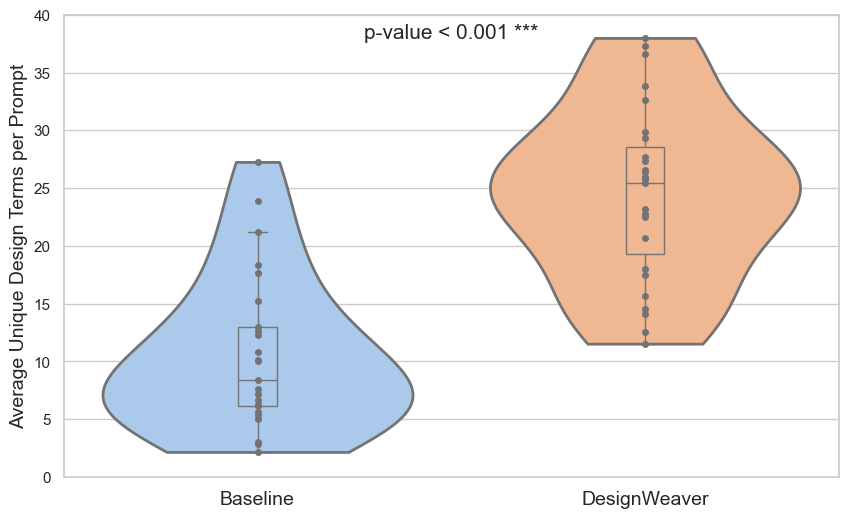}
    \caption{DesignWeaver participants averaged more unique design terms per prompt than the Baseline.}
    \Description{A violin plot showing the distribution of the average number of unique design terms per prompt for Baseline and DesignWeaver groups. The DesignWeaver group has a higher median and a higher range overall.}
    \label{fig:findings_avg_unique_design_terms}
\end{figure}

\begin{figure}[htbp]
    \centering
    \includegraphics[width=\linewidth]{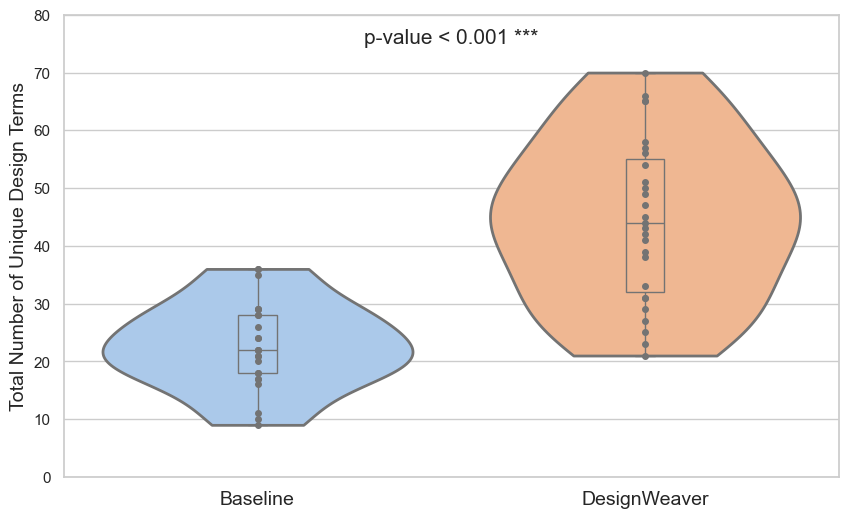}
    \caption{DesignWeaver participants had a higher total number of unique design terms per session than the Baseline.}
    \Description{A violin plot showing the distribution of the total unique tags per session for Baseline and DesignWeaver groups. Again, the DesignWeaver group exhibits a higher median.}
    \label{fig:findings_total_unique_design_terms}
\end{figure}

In addition to generating longer prompts, participants in the DesignWeaver group employed a broader range of unique design terms. \autoref{fig:findings_avg_unique_design_terms} shows, on average, participants in the \toolname{} group used 24.48 unique terms per prompt (SD = 7.47), compared to 10.59 unique terms in the Baseline group (SD = 6.72). Similarly, \autoref{fig:findings_total_unique_design_terms} shows that the total number of unique tags per session was also higher in the \toolname{} group (M = 44.44, SD = 14.07) than in the Baseline group (M = 22.72, SD = 7.50), with significant differences confirmed by the analysis ($U = 61.0, p < 0.001$ \& $U = 56.5, p < 0.001$).

\begin{figure}[htp]
    \centering
    \includegraphics[width=\linewidth]{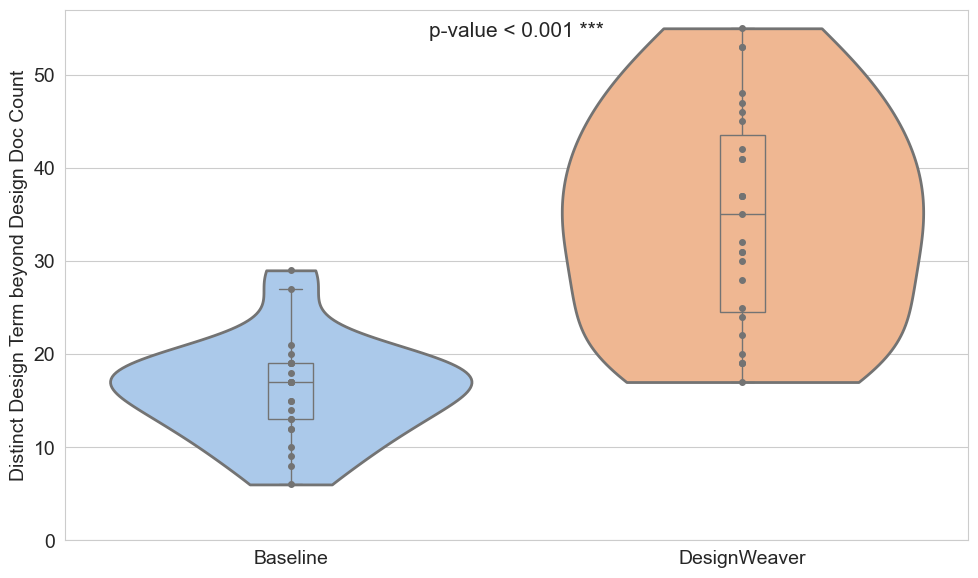}
    \caption{DesignWeaver participants come up with more distinct design terms than the Baseline.}
    \label{fig:findings_prompt_design_term_violin_plots_unique}
    \Description{A violin plot comparing the distribution of distinct design terms for Baseline vs. DesignWeaver. The DesignWeaver group has a higher median and a broader range.}
\end{figure}

\begin{figure}[htp]
    \centering
    \includegraphics[width=\linewidth]{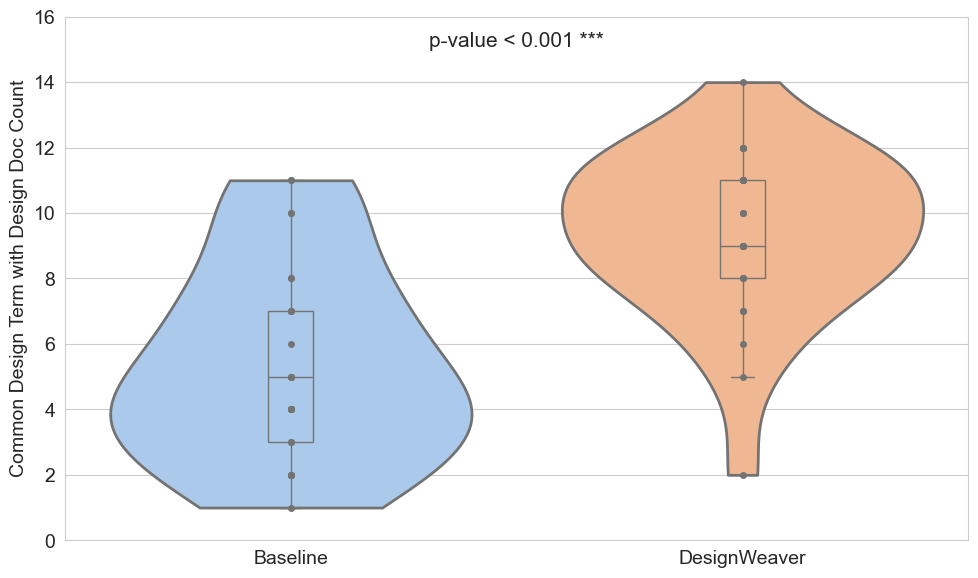}
    \caption{DesignWeaver participants adapted more common design terms than the Baseline.}
    \label{fig:findings_prompt_design_term_violin_plots_common}
    \Description{A violin plot comparing the distribution of common design terms for Baseline vs. DesignWeaver. Again, the DesignWeaver group shows a higher median usage.}
\end{figure}

To assess the impact of dimensional scaffolding on design vocabulary development, we compared, between the two groups, how many of the design terms were directly adapted from the Design document and how many were developed by the participants themselves while using the tool. Participants using \toolname{} both thought of more distinct design terms ($M = 34.6, SD = 11.7$) and common design terms ($M = 9.4, SD = 2.5$) from the Design document than those in the Baseline group, who had distinct terms ($M = 16.2, SD = 5.3$) and common design terms ($M = 5.4, SD = 3.0$), as shown in \autoref{fig:findings_prompt_design_term_violin_plots_unique} and \autoref{fig:findings_prompt_design_term_violin_plots_common}. Statistically significant differences were observed in the adoption rates for distinct terms ($U = 635.0, p < 0.001$) and common design terms ($U = 566.0, p < 0.001$), demonstrating that dimensional scaffolding facilitated not only the use of terms from the Design document but also encouraged participants to learn and adopt new design terms.

\subsubsection{Participants Adapted Diverse Prompt Strategies Afforded by \toolname{}}
\label{finding6.1.2:prompt_strategy}

\begin{figure*}[htbp]
    \centering
    \includegraphics[width=\textwidth]{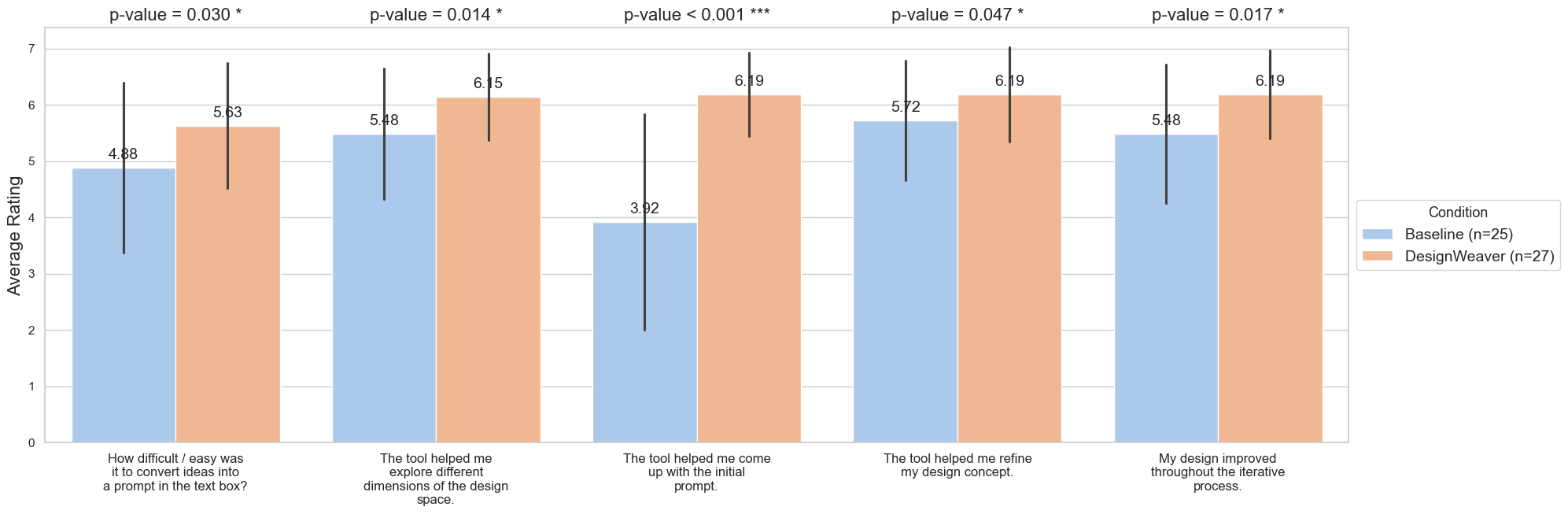}
    \caption{Participants rated DesignWeaver higher than the Baseline on ease of idea-to-prompt conversion, design space exploration, prompt generation, concept refinement, and iterative design improvement.}
    \label{fig:finding_survey_average_ratings_comparison}
    \Description{This figure compares the average ratings given by participants in the Baseline (n=25) and DesignWeaver (n=27) groups across multiple survey questions. The p-values above the bars show statistical significance for some of the questions. The x-axis lists the survey questions, and the y-axis represents the average ratings on a Likert scale. Statistically significant results are marked, with differences favoring the DesignWeaver condition in several categories.}
\end{figure*}

Participants using \toolname{} employed multiple strategies during prompt crafting, integrating dimension scaffolding to guide their design process. Initially, all participants (27/27) started with automated prompts generated from selecting tags based on the Design document. Throughout the process, they employed three main strategies: 18 out of 27 used default tags from the discussion panel, 5 out of 27 created custom dimensions and tags, and 12 out of 27 manually edited their prompts. Additionally, 6 out of 27 participants used the information button to map images back to dimension tags, gaining insights into how specific elements influenced the design. Notably, 10 out of 27 participants switched between these strategies. For example, P24 began with default tags such as ``Minimalist,'' ``Eco-friendly,'' `` Ergonomic,'' and ``Scratch-resistant.'' After reviewing the initial images, they added new dimensions and tags from the recommended list. As they continued, P24 customized tags using terms from the design document, like ``stain-resistant,'' and further refined their prompts by manually adding specific details such as ``Width is 18 inches and depth is 16 inches.'' Adopting this expanded range of approaches, participants using \toolname{} found it significantly easier ($U = 236.5, p = 0.030$) to convert ideas into prompts, as reflected in their higher ratings ($M = 5.63, SD = 1.11$) compared to the Baseline group ($M = 4.88, SD = 1.51$) shown in \autoref{fig:finding_survey_average_ratings_comparison}. 

In contrast, all participants (25/25) in the Baseline group relied heavily on vocabulary from design documents and adjusted their prompts iteratively based on visual feedback from generated images. For straightforward requirements like ``Solid oak'', participants easily incorporated terms directly from the design documents. For example, P2 stated, ``\textit{I mainly just copied and pasted from the document and... included those into the prompt.}" When dealing with more complex requirements, such as ensuring comfort in long conversations, 24 out of 25 participants started with general terms and refined them based on feedback. As P21 explained, they began with ``\textit{neutral colored with a cushion seat,}'' then added features such as wood material and color accents based on how images aligned with their vision, iteratively refining to achieve a closer match. Such reliance on image feedback was reported to lead to participants' frustration when images failed to improve; 9 of the 25 participants reported this experience. For instance, P46 noted, \begin{quote}``\textit{while I'm … adding more information or … keywords to it, but sometimes … some keywords later, it doesn't change that much. It just gives me some new pictures, but I can't really … tell where… there's a big change.}''\end{quote}

\subsection{Impact of Dimensional Scaffolding on Generated Artifacts}
\label{finding6.2:image_novelty}
Using the ViT-B/32 CLIP model \cite{radford2021learning}, we calculated both image and prompt similarities for the Baseline and \toolname{} conditions. Results show that:

\begin{figure}[htbp]
    \centering
    \includegraphics[width=\linewidth]{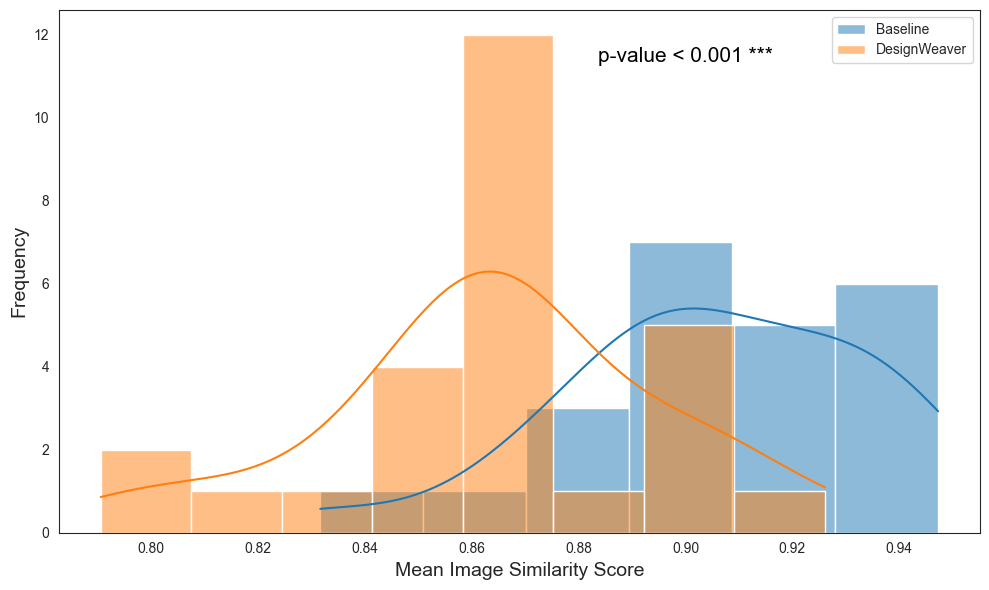}
    \caption{DesignWeaver participants created semantically more diverse images than the Baseline.}
    \label{fig:finding_image_similarity_scores_distribution}
    \Description{Histogram comparing the mean image similarity scores for the Baseline (blue) and DesignWeaver (orange) groups. The p-values from the Mann-Whitney U Test indicate that DesignWeaver has significantly higher image similarity scores.}
\end{figure}

\begin{figure}[htbp]
    \centering
    \includegraphics[width=\linewidth]{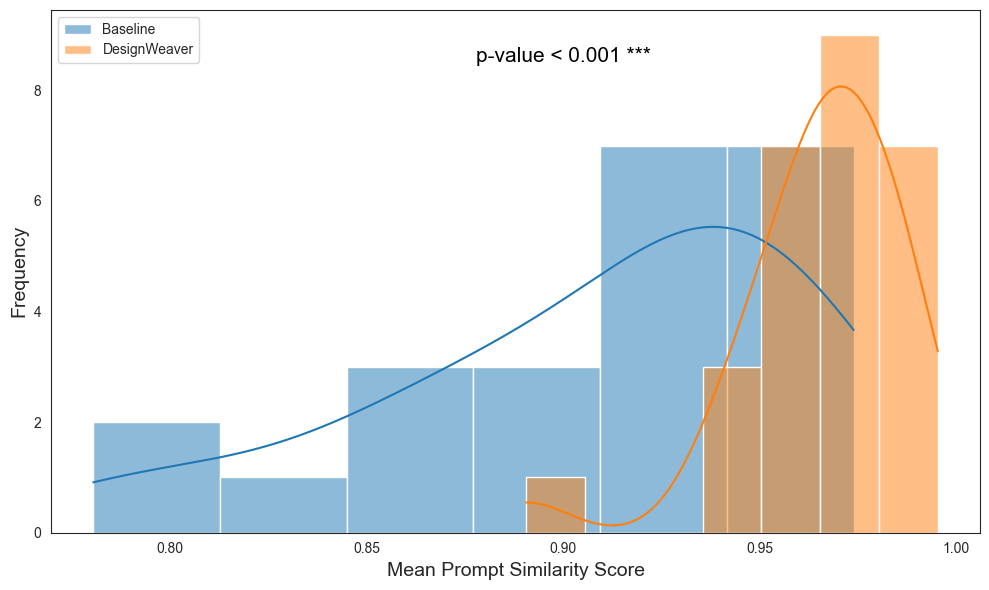}
    \caption{DesignWeaver participants created semantically more similar prompts than the Baseline.}
    \label{fig:finding_prompt_similarity_scores_distribution}
    \Description{Histogram comparing the mean prompt similarity scores for the Baseline (blue) and DesignWeaver (orange) groups. The p-values from the Mann-Whitney U Test indicate that DesignWeaver has significantly higher prompt similarity scores.}
\end{figure}


Both image and prompt similarity scores were significantly different from each other group (p<0.001). In image similarity, Baseline has a mean similarity of 0.903 (SD=0.015), and \toolname{} has a similarity of 0.863 (SD=0.024), which yields a statistically significant difference ($U=521.0, p<0.001$) (see \autoref{fig:finding_image_similarity_scores_distribution}). In prompt similarity, Baseline has a mean similarity of 0.916 (SD=0.035), and \toolname{} has a similarity of 0.964 (SD=0.022), which also yields a statistically significant difference ($U=78.0, p<0.001$) (see \autoref{fig:finding_prompt_similarity_scores_distribution}). We included more detailed analysis in Appendix~\ref{appD:sd}.

\begin{figure}[htbp]
    \centering
    \includegraphics[width=\linewidth]{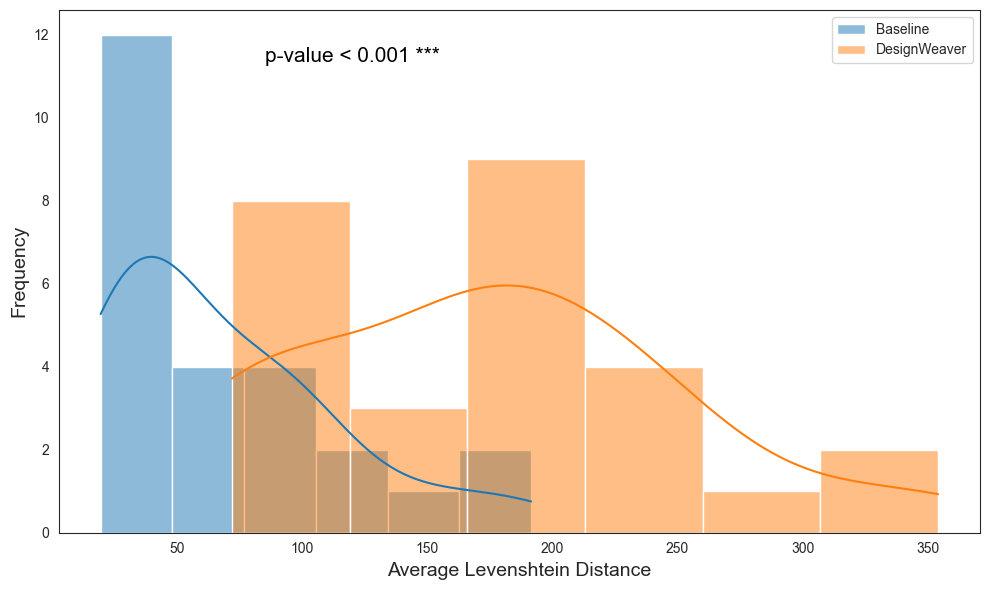}
    \caption{DesignWeaver group has larger Levenshtein text edit distance in prompts across iterations than the Baseline.}
    \label{fig:finding_levenshtein_distances_distribution}
    \Description{This figure presents the distribution of average Levenshtein distances for Baseline in blue and DesignWeaver in orange. The x-axis represents the average Levenshtein distance (a measure of textual difference between prompts), and the y-axis represents the frequency of occurrences. The density curves highlight that DesignWeaver tends to have higher average Levenshtein distances than the Baseline, indicating more variation in the textual evolution of prompts in the DesignWeaver group.}
\end{figure}

To further explore why that semantically similar prompts in \toolname{} yield more semantically diverse image outcomes, we performed a simpler prompt difference comparison using the Levenshtein edit distance to address potential concerns. The results in \autoref{fig:finding_levenshtein_distances_distribution} indicate that users in the \toolname{} condition ($M=174.49, SD=74.83$) made significantly more prompt modifications compared to the Baseline Condition ($M=68.21, SD=44.94$). A Mann-Whitney U Test produced a U statistic of 76.0 and a p-value less than 0.001, confirming the robustness of our findings.

\begin{figure*}[htbp]
    \centering
    \includegraphics[width=\textwidth]{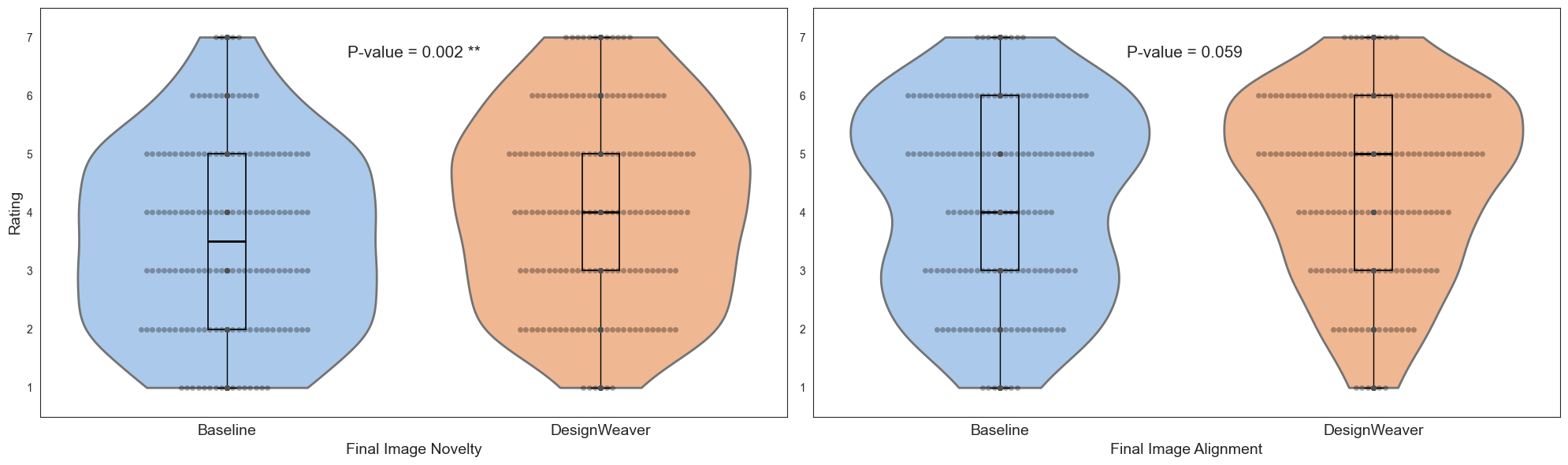}
    \caption{Average expert ratings on the novelty and alignment with the client's design brief of the participants' final design. DesignWeaver participants created designs with significantly higher ratings on novelty but not alignment compared to baseline participants.}
    \Description{Two side-by-side violin plots comparing expert ratings on Novelty and Value Alignment between Baseline and DesignWeaver groups. The left plot displays a significant difference, while the right plot does not.}
    \label{fig:findings_expert_ratings}
\end{figure*}

\begin{figure}[htbp]
    \centering
    \includegraphics[width=\linewidth]{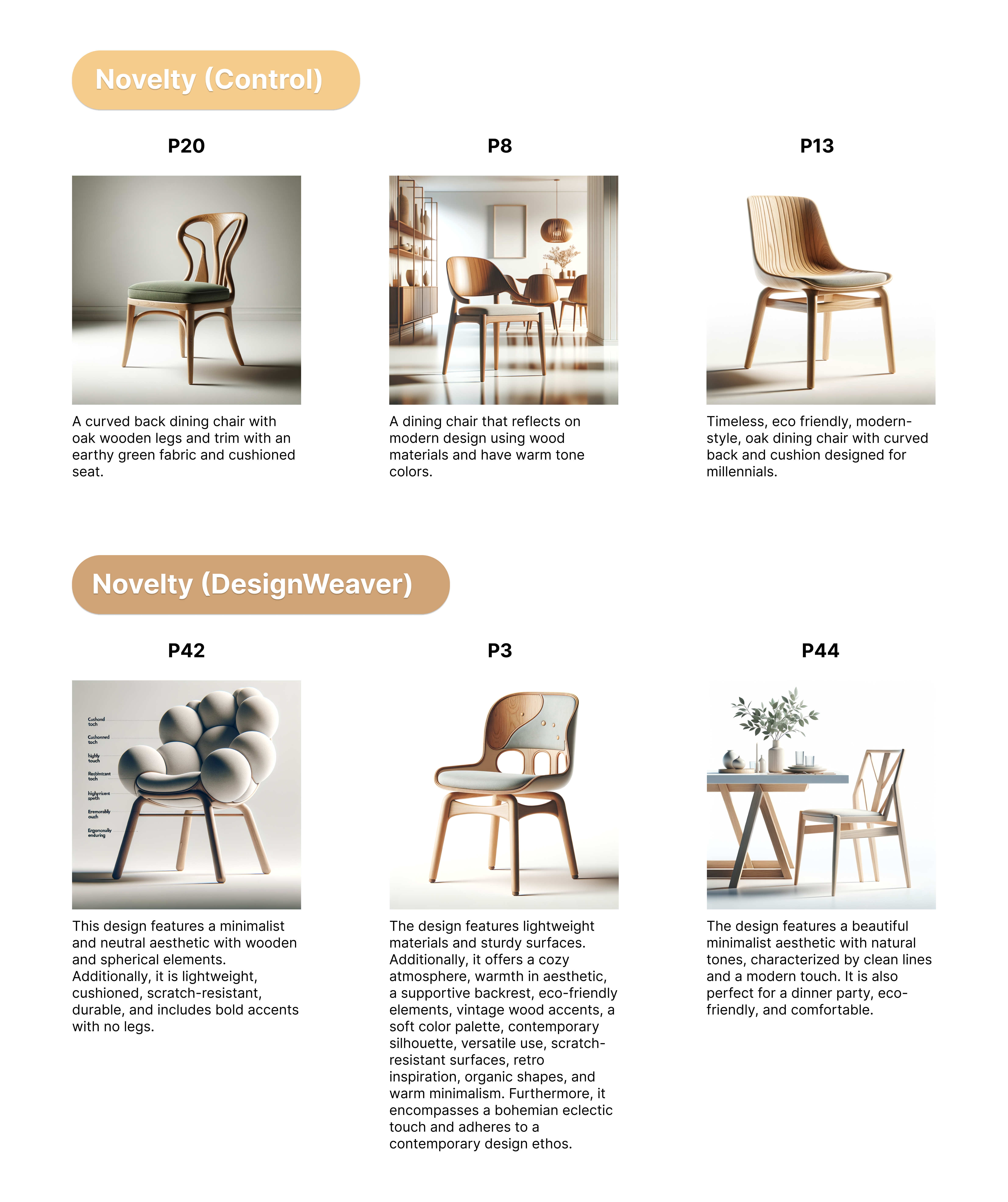}
    \caption{Top 3 expert rated chair on innovation}
    \label{fig:finding_novelty_gallery}
    \Description{The figure presents two sets of designs: the baseline group (top) and the DesignWeaver group (bottom). The baseline group focuses on minimalist designs with moderate creative elements, whereas the DesignWeaver group showcases more distinctive and novel designs that integrate sustainability, creative material choices, and innovative forms. Each design is labeled with its unique identifier.}
\end{figure}

Lastly, we compared the experts' rating data we gathered on novelty and value alignment for the final images picked by all the participants. We found that \toolname{} condition was rated significantly higher on image novelty ($M=4.09, SD=1.63$) compared to the Baseline group ($M=3.54, SD=1.60$), with a statistically significant difference ($U=9899.0, p=0.002$), as illustrated in \autoref{fig:findings_expert_ratings} (left). Analysis of the alignment with design requirements didn't show a significant difference between \toolname{} condition ($M=4.5, SD=1.656$) and Baseline condition ($M=4.2, SD=1.67$) with a statistical significance at 0.059 ($U = 10925.5$), shown in \autoref{fig:findings_expert_ratings} (right). \autoref{fig:finding_novelty_gallery} showcases the top 3 chairs rated by experts for their novelty.

\subsection{Users Perceived Support and Challenges}
\label{finding6.3:perception}

\subsubsection{ \toolname{} Participants Adapted New Design Terms}
One of the primary advantages of \toolname{} reported by participants was its ability to help participants adopt new design terminology. 
\toolname{} was perceived to significantly aid in adapting new design terms, as evidenced by significantly higher ($U = 225.0, p = 0.0139$) ratings from participants using \toolname{} ($M = 6.15, SD = 0.77$) compared to the Baseline group ($M = 5.48, SD = 1.16$) for exploring different design dimensions (see \autoref{fig:finding_survey_average_ratings_comparison}). This adaptability was noticeable from the initial stages of the design process, where \toolname{} generated images that closely aligned with participants’ early visions. Participants found \toolname{} highly effective in helping create initial prompts ($M = 6.19, SD = 0.74$), significantly outperforming ($U = 115.0, p < 0.001$) the Baseline group ($M = 3.92, SD = 1.91$). For instance, P7 remarked, “\textit{it immediately generated something very similar to the mood board,}” while P11 noted that it “\textit{helped generate a sort of Baseline on what chairs I could generate.}” Throughout the design process, the tags listed on the panel were instrumental in broadening participants' design exploration, with 8 participants appreciating how these tags introduced new dimensions and terminology they hadn't previously considered. P20 highlighted, “\textit{there were lots of different dimensions that I wouldn’t typically think about for chairs. So that was like new information.}” Similarly, P13 valued the exposure to terms like “ergonomic” and “sustainability,” which allowed them to refine and expand their design ideas, enhancing their creative versatility.

\subsubsection{\toolname{} Participants Gained Greater Control Over the Design Process}
Participants also felt that \toolname{} enhanced their control over the design process, enabling them to manage their designs more effectively through tag selection. 8 of the 27 participants mentioned that tag selection helped them efficiently navigate and refine their designs. P1 described their experience: ``\textit{I was able to change the chair and select different [options] that fit the design... and go back if I wanted to add or change... tags.}'' Among these 8 participants, six highlighted the information button as crucial for staying informed about their current design state and understanding how each modification impacted their design within specific dimensions. P30 explained, ``\textit{Whenever I got info from the chairs that I liked, it was helpful to see what wording or characteristics I should remember.}'' Participants using \toolname{} reported a greater sense of control, reflected in significantly higher ratings for refining their design concepts ($M = 6.19, SD = 0.74$) compared to the Baseline group ($M = 5.72, SD = 1.06$; $U = 251.5, p = 0.047$) as shown in \autoref{fig:finding_survey_average_ratings_comparison}. They also felt their designs improved more throughout the iterative process ($M = 6.19, SD = 0.79$) than those in the Baseline group ($M = 5.48, SD = 1.23$; $U = 226.5, p = 0.017$), further supporting the tool’s effectiveness.

\subsubsection{\toolname{} Participants Faced Challenges with Over-Tagging and Generating Specific Designs}
During interviews, participants also reported challenges while using \toolname{}. A notable issue was that choosing too many tags (over-tagging) led to confusion over the design process. Several participants (5 out of 27) reported that over-tagging led to confusion and a sense of losing baseline over the design process. P11 mentioned, ``\textit{if you choose too many options, it'll get a bit too constricted,}'' reflecting a perception that too many tags can make the system overly restrictive. P1 similarly felt that ``\textit{at the end, it was kind of already having too many tags, and it was kind of just showing me like a little off to what I needed.}'' Participants also expressed concerns about the tool's responsiveness to tag changes. For example, P17 stated, ``\textit{I feel like when I change the culture design tab, not much changed,}'' suggesting that tag adjustments did not always produce the expected variations.

Furthermore, participants reported difficulties in generating designs with specific or detailed characteristics. P47 shared their experience: ``\textit{There were many times when I was clicking around trying to get more variety in an image… but it didn't really do that. Towards the end, I started getting closer, but it wasn't what the client wanted. The client wanted a curved back [on the chairs], but instead of giving me curved backs as in the beginning, it started giving me straight backs.}'' Similarly, P18 found it challenging to generate specific colors, saying, ``\textit{I wanted the colors not to be just tan and white, but they were only giving me tan and white. I didn't know how specific I could be or if I could just click certain colors.}'' P5 also described the increased effort needed as designs became more complex, stating, ``\textit{as we got further on into the more nitty, gritty aspects, then it was more on me,}'' reflecting the perceived effort needed to refine the design as complexity increased.

With perceived challenges and support from \toolname{}, participants using the tool reported similar satisfaction levels with the design images compared to the Baseline group, both in terms of quality and alignment with their expectations. There was no significant statistical difference between the Baseline and \toolname{} conditions for satisfaction with the generated images ($M = 5.56, SD = 1.04; M = 5.59, SD = 0.63; p = 0.5790$) and for alignment of the images with expectations ($M = 5.44, SD = 0.87; M = 5.63, SD = 0.84; p = 0.1314$).

\section{DISCUSSION}
Based on a formative study with experts, we prototyped a creativity-support tool to explore how dimensional scaffolding can bridge the gap between novice designers’ capabilities and the complex demands of product design using generative AI. At its core, this research proposes and evaluates \toolname{}, an interaction paradigm that surfaces key product design dimensions from user-curated images into a selectable palette to enhance text prompts for text-to-image generation models.

The formative study revealed insights into client-designer dynamics and novice-expert gaps in design exploration. Designers frequently rely on visual aids and multiple iterations to communicate ideas and align with clients’ preferences, emphasizing the importance of managing trade-offs and balancing creative goals with practical constraints. These findings underscored the need for an interface that emphasizes \textit{visual feedback} and \textit{structured exploration}, especially for users without domain-specific knowledge. These insights informed \toolname{} ’s "dimensional scaffolding" paradigm, which we hypothesized would allow novices to discover domain language and more fluently explore design spaces typically reserved for experts. In the study, dimensional scaffolding in \toolname{} enhanced textual descriptions and supported iterative design processes, enabling novice designers to produce semantically diverse, novel outcomes while fostering greater self-efficacy through perceived control and improvement. Below, we delve into the interaction paradigm of \toolname{}, leveraging relevant theories to explain the outcomes and implications.

\subsection{Why did \toolname{} yield more diverse image outputs?}
Participants using \toolname{} produced longer prompts with richer vocabularies, contributing to images of greater diversity. By surfacing key design dimensions and associated vocabulary, \toolname{} encouraged users to craft detailed textual descriptions that articulated nuanced design ideas. These longer and richer prompts provided generative models with the necessary context to produce varied and novel design outputs.

While the \toolname{} creations were more diverse, participants’ iterative prompting exhibited higher semantic similarity than the baseline participants. Those minor edits tended to be meaningful refinements, leveraging the dimensional scaffolding framework. This suggests that while participants refined their designs, they preserved a consistent conceptual design. At the same time, the higher Levenshtein edit distances across iterations indicate that participants made substantial textual edits to their prompts. These edits introduced nuanced semantic adjustments, which allowed participants to explore new design possibilities while staying grounded in their core design goals. This coherence aligns with cognitive theories of design iteration in which minor, deliberate modifications within a structured framework lead to richer, more meaningful creative outputs \cite{karimi2019relating, davis2024fashioning}. Such iterative exploration and diversity in outputs reflect the value of generating and sharing multiple prototypes \cite{dow2011prototyping, hartmann2008design, kim2022mixplorer}, which promotes deeper creative exploration and more well-rounded design solutions.

\subsection{To what extent did \toolname{} participants understand the design space?}

Participants using \toolname{} demonstrated a better ability to navigate the design space, as evidenced by their significantly higher ratings compared to the baseline group for exploring different design dimensions and their significantly lower reported difficulty in converting ideas into text-based prompts.

This enhanced exploration is attributed to the dual feedback loop facilitated by \toolname{} ’s interaction paradigm (see section~\ref{finding6.3:perception}). Dimensional scaffolding allowed participants to map textual inputs to visual outputs, offering immediate feedback on how specific design terms influenced generated images \cite{yen2024give}. Simultaneously, the inspect feature supported the reverse process by linking visual outputs to associated design tags, making implicit design elements explicit. This iterative interplay between exploration and interpretation enabled participants to refine their mental models of the design space and articulate their creative intentions more effectively. Such scaffolding is particularly valuable in design contexts where visual thinking dominates \cite{xu2024jamplate,heyman2024supermind, dhillon2024shaping}, but articulating visual concepts verbally can often be challenging \cite{zamfirescu2023johnny}.

\subsection{How did \toolname{} affect the co-creation process?}
Participants using \toolname{} reported significantly higher ratings for “My design improved throughout the iterative process,” reflecting their positive perception of the system’s ability to support iterative refinement. Expert evaluations further confirmed these results, with significantly higher ratings for the novelty and alignment with the client’s design brief in participants’ final designs than in the baseline group. These outcomes demonstrate \toolname{}’s effectiveness in fostering creative and contextually relevant designs.

This improvement can be partly attributed to the mechanisms discussed above, where participants were able to explore a broader range of design possibilities and simultaneously deepen their understanding of the design space. The system’s bi-directional feedback loop and dimensional scaffolding enabled participants to refine the prompt dynamically, connecting design dimensions with visual outputs \cite{green1989cognitive, davis2018scaffolding}. By engaging with these iterative processes, participants can refine their prompts effectively and articulate nuanced design ideas, balancing creativity with alignment with client goals.

However, despite these positive outcomes, participants did not express significantly higher satisfaction with the generated images. This discrepancy likely stems from heightened expectations created by the system’s structured and adaptable framework. By enabling iterative refinement and supporting creative exploration, \toolname{} may have amplified participants’ expectations for the specificity and quality of the generated outputs \cite{amershi2014power, zhou2024understanding}. While effective at capturing broad design ideas, current generative models often struggle to deliver the detailed and precise characteristics participants sought. This highlights a critical gap between user empowerment and the technical limitations of existing generative AI, suggesting a need for advancements that better align outputs with users’ elevated expectations \cite{lee2024holistic}. 

\subsection{Limitations}

While the study demonstrates the potential of dimensional scaffolding in enhancing novice designers’ creative workflows, several limitations should be acknowledged. 

Firstly, although participants using \toolname{} reported an increased sense of control with structured prompts, we also received complaints when the generated images did not meet their expectations, reflecting the widening execution and evaluation gulfs \cite{subramonyam2024bridging}. This underscores a limitation in current text-to-image models \cite{lee2024holistic}, which struggle to consistently render detailed prompts accurately. As these models evolve, they may better satisfy user needs by accurately interpreting nuanced design specifications. Nonetheless, novices' mastery of domain-specific design language remains crucial.

Secondly, the study was confined to the domain of product design, specifically focusing on chair designs. Future studies in different domains, such as fashion \cite{jeon2021fashionq}, UI design \cite{vaithilingam2024dynavis}, or architecture \cite{aseniero2024experiential, zhang2023generative}, will help strengthen insights on scaffolding novice interactions with AI models. 

Lastly, while dimensional scaffolding provides a helpful structure for thinking about well-established aspects of a design space, it may also constrain creativity by limiting users to primary dimensions and tags. Finding the optimal balance between offering helpful guidance and allowing creative flexibility is challenging \cite{chi2014nature}, especially for users with varying levels of expertise \cite{cross2004expertise, ericsson2009development}. The system’s current design might not fully accommodate the needs of more experienced designers who may prefer less restrictive tools.

\subsection{Future Work}

Building on our findings, several future research and development directions emerge. Enhancing designers’ control over generated outputs and addressing system limitations are key to improving the creative process. Features like image- \cite{zhang2023adding} or sketch-conditioned generation \cite{chang2020workflow, sarukkai2024block}, in-painting \cite{lugmayr2022repaint}, and direct manipulation \cite{pan2023drag} could give users greater control by allowing them to retain and modify specific design geometries during exploration \cite{brade2023promptify}. Additionally, well-timed feedback \cite{yen2024give}, dynamic scaffolding \cite{xu2024jamplate, suh2024luminate}, analogical thinking \cite{lin2025inkspire, kang2024biospark}, and advanced agentic generative models \cite{park2023generative, shaikh2024rehearsal} could bridge the gap between user expectations and outputs. These enhancements would offer more accurate and consistent translations of detailed prompts into visual designs, making the creative process more intuitive, satisfying, and productive.

Second, investigating how to tailor the scaffolding to suit different levels of expertise could help find the optimal balance between providing guidance and allowing creative freedom, ensuring that both novices and experienced designers benefit from the approach. 
While predefined dimensions offer essential guidance for novices, the ideal level of customization for experienced users remains an open question. Future research could compare tools prioritizing strict predefined structures with those offering greater customization to explore how different scaffolding strategies impact user satisfaction, creativity, and outcomes. Studying tools with varying degrees of dimensional scaffolding across different expertise levels could reveal the optimal balance between structure and flexibility.

Furthermore, improving the visualization and management of design metadata is essential, especially when dealing with many generated designs. Exploring alternative visualization methods—such as hierarchical trees, interactive maps, or layered sheets—could help users better navigate and understand the design space \cite{jiang2023graphologue, wootton2024charting}. Enhanced visualization tools could enable designers to track their iterative changes more effectively, compare different design variations, and make more informed decisions throughout the creative process.

Finally, adapting the system for collaborative environments presents an intriguing opportunity. Future work could explore how dimensional scaffolding might support collaborative design workflows, enabling multiple designers or designer-client teams to co-create with AI assistance. Investigating collaborative setups could provide deeper insights into supporting diverse creative processes and enhancing team-based design projects \cite{bodker2000creativity}. Future work could explore the potential for LLMs to facilitate communication and idea sharing among collaborators \cite{he2024ai, heyman2024supermind}, to support convergence processes among various stakeholders \cite{rayan2024exploring}, or to enable consensus-building \cite{liu2018consensus}.

\section{CONCLUSION}
In this work, we introduced \toolname{}, a generative AI system that empowers novice designers through dimensional scaffolding, encouraging broader design exploration and more nuanced prompt creation. Our findings demonstrate that \toolname{} led to significantly longer and more nuanced prompts, which produced more semantically diverse image output. It also better enabled participants to discover and adopt new design dimensions. Compared to a baseline, \toolname{} users produced more unique design terms and exhibited greater confidence in navigating complex design challenges. They engaged in richer creative processes, integrating both more predefined and custom tags to refine their designs iteratively. This suggests that dimensional scaffolding can effectively bridge the gap between novice and expert design strategies, offering a powerful tool for enhancing creative workflows.
\begin{acks}
We would like to thank Prof. Scott Klemmer for his insightful advice during the early stages of this project while it was still a course research project. We also thank our TA for the course, Jaidev Shriram, for his valuable input and feedback. Special thanks to Yung-Wei Yang for her assistance in writing some early-stage code for the front end and for bringing valuable ideas to our group meetings. We are also grateful to Emilia Rosselli-del-turco and Peiling Jiang for their thoughtful insights on the project. Thanks also to Jude Rayan, Lu Sun, Jeongeon Park, Tone Xu, and Bryan Min for their tips on good practices for data analysis, visualization, chart \& diagram creation, and paper writing. Lastly, we thank Jude’s dog, Idli, for emotionally supporting the design lab during the CHI deadline crunch. This research was funded by NSF grant \#2009003.

We would also like to thank the CHI ACs and the anonymous reviewers for their efforts in reviewing this paper and for their constructive feedback, which greatly improved the paper's quality.

We produced all the original writing in this manuscript and used ChatGPT to assist in shortening and modifying the text. Perplexity was used for literature searches, and GitHub Copilot and ChatGPT were employed for coding, data analysis, and visualization.
\end{acks}

\newpage
\bibliographystyle{ACM-Reference-Format}

\appendix

\section{PROMPTS USED IN THE LLM PIPELINES}
\label{appA:prompts}
\subsection{Document Digestion}
\label{appA1:prompts_doc_digest}
\begin{itemize}
    \item \textbf{prompt}: "Get the three most important design dimensions from the requirement doc. For each dimension, generate 3-5 tags. Be concise."
\end{itemize}

\subsection{Prompt Generation \& Update}
\label{appA2:prompts_prompt_gen_and_update}
\begin{itemize}
    \item \textbf{system\_message}: "You are a design generalist that converts design tags and weights into descriptive prompts. Your task is to update the prompt according to the given old and new tags comparison and their corresponding weights, making sure to remove any references to tags that have been removed or neutralized (weight = 0). Preserve as much of the original prompt as possible, but reflect all tag changes accurately."
    \item \textbf{user\_message}: "Create a product rendering of a dining room chair that stands out prominently against a white background. Update the old prompt by comparing the old and new tags and weights pairs. Any tags with a weight of 0 should be removed from the prompt. Any tags with a weight of 1 should be included in the prompt.
    \newline
    New Tags: \$\{JSON.stringify(transformedNewTags)\}
    \newline
    Old Tags: \$\{JSON.stringify(transformedOldTags)\}
    \newline
    Old Prompt: "\$\{promptText\}"
    \newline
    Just return the prompt itself. Use complete sentences to describe the design."
    \item \textbf{model}: "gpt-4o"
    \item \textbf{messages}: [system\_message, user\_message]
\end{itemize}

\subsection{Image Generation}
\label{appA3:prompts_img_gen}
\begin{itemize}
    \item \textbf{model}: "dall-e-3"
    \item \textbf{quality}: hd
    \item \textbf{num}: 3
    \item \textbf{size}: "1024x1024"
    \item \textbf{quality}: "standard"
\end{itemize}

\subsection{Tag Extraction}
\label{appA4:prompts_tag_extraction}
\begin{itemize}
    \item \textbf{system\_message}: "You are a creative and helpful designer who assists in identifying and categorizing aesthetic dimensions of product designs. The response should be format like: \{newtags:[\{'name':'Dimension Name', 'tags':['tag1', 'tag2', 'tag3' ... \}]\}"
    \item \textbf{text\_prompt}: "What relevant aesthetic dimensions and design element tags are in this image? Reference the existing tags, think outside the box, and include all relevant dimensions. On top of the old tags, generate 1-5 new tags that either append to existing design dimensions or create new dimensions while avoiding creating similar dimensions to the old ones. Provide the output in a JSON format." + JSON.stringify(JSON\_prompt);
    \item user\_message: [
    \newline
    \{type: "text", text: text\_prompt\}, 
    \newline
    \{type: "image\_url", image\_url: \{url: url,detail: "low",\}
    \newline
    \}]
    \item \textbf{model}: "gpt-4o-mini"
    \item \textbf{response\_format}: \{ type: "json\_object" \}
    \item \textbf{messages}: [system\_message, user\_message]
\end{itemize}

\subsection{Tag Recommendation}
\label{appA5:prompts_tag_recommendation}

\begin{itemize}
    \item \textbf{system\_message}: "You are a helpful assistant that provides concise five distinct design recommendations based on existing design tags for dining room chair design."
    \item \textbf{user\_message}: "Based on the current design tags in the category '\$\{currentCategory.name\}', suggest five new distinct design options. Please provide a simple list of options separated by commas and nothing else. Don't add numbers or bullet points."
    \item \textbf{model}: "gpt-4o-mini"
    \item \textbf{messages}: [system\_message, user\_message]
\end{itemize}

\subsection{Dimension Recommendation}
\label{appA6:prompts_dimension_extraction}
\begin{itemize}
    \item \textbf{system\_message}: "You are a helpful assistant that provides concise design recommendations based on existing design dimensions for dining room chair design."
    \item \textbf{user\_message}: "Based on the current design dimensions: [\$\{categoriesList\}], suggest 5 new distinct dimensions. Please provide a simple list of dimensions separated by commas and nothing else. Don't add numbers or bullet points."
    \item \textbf{model}: "gpt-4o-mini"
    \item \textbf{messages}: [system\_message, user\_message]
\end{itemize}

\section{USER STUDY RELATED MATERIAL}
\subsection{Design Document}
\label{appB:dd}
\subsubsection{Design requirement}

(Email from our client David)

Dear Designer,

I hope this message finds you in great spirits. As an architect deeply passionate about modern and sustainable design, I'm reaching out with a project that's very close to my heart. I'm in the process of bringing a vision to life—a set of 10 dining room chairs that's not just a piece of furniture but a statement of my lifestyle and values.

\textbf{Here's What I Envision:}
I'm drawn to contemporary style and embrace minimalism with open arms. My home is a testament to my love for clean lines and functional aesthetics, and I seek to extend this philosophy to this new chair design. I imagine it in neutral tones, with customization options allowing for the occasional bold accent. The materials, of course, need to echo my commitment to sustainability—a combination of natural wood and high-quality, eco-friendly fabrics would be ideal.

\textbf{Design Specifications:}
\begin{itemize}
    \item \textbf{Dimensions}: A height of 36 inches, with a seat height of 18 inches, seems perfect. The width and depth should be about 18 and 16 inches, respectively, providing ample space without compromising the minimalist design.
    \item \textbf{Comfort and Ergonomics}: Given my love for hosting dinner parties, the chair must offer comfort for long conversations. A curved backrest for lumbar support and a cushioned seat are essential.
    \item \textbf{Materials}: Solid oak or a similar hardwood for the frame would complement my home's aesthetic, paired with durable and stain-resistant fabric sourced responsibly.
\end{itemize}

\textbf{Functionality Needs:}
The chair should be lightweight, making it easy to move around, yet sturdy enough to withstand the joyous chaos of my gatherings. I also need to ensure the feet are kind to my flooring—no scratches are welcome.

\textbf{Sustainability and Quality:}
I value furniture that reflects my dedication to sustainable living and modern design. It should be durable and sourced from eco-friendly suppliers, aligning with my values and lifestyle.

\textbf{Budget and Timeline:}
I've allocated \$1000 to \$1500 per unit for this project, aiming to balance impeccable quality and affordability. Ideally, the design would be finalized within 30 days, with a prototype ready for review 60 days later and production commencing 90 days after prototype approval.

\textbf{Additional Considerations:}
Ease of assembly and eco-friendly packaging that ensures safe transport without compromising our planet's health is crucial.

I'm thrilled at the prospect of working together to create a piece that serves its purpose and does so with style and conscience. Your talent in design and understanding of functional aesthetics make you the perfect partner for this venture.

Looking forward to your thoughts and ideas.

Best regards,
David Thompson

\subsubsection{Client’s Persona}
\begin{itemize}
    \item David Thompson
    \item Age: 35
    \item Occupation: Architect
    \item Location: San Francisco, California
    \item Marital Status: Married
    \item Interests:
    \begin{itemize}
        \item Passionate about modern and sustainable architecture.
        \item Enjoys reading about interior design and home renovation.
        \item Loves outdoor activities like hiking and cycling.
    \end{itemize}
    \item Lifestyle:
    \begin{itemize}
        \item Lives in a well-designed, modern home.
        \item Prefers a minimalist and functional aesthetic.
        \item Often hosts dinner parties and small gatherings.
        \item Has a few dogs of varying ages
    \end{itemize}
    \item Purchase Motivations:
    \begin{itemize}
        \item Seeking furniture that reflects his taste for modern design.
        \item Values sustainability and eco-friendly materials.
        \item Wants durable, high-quality furniture that can withstand regular use.
    \end{itemize}
    \item Functional Needs:
    \begin{itemize}
        \item Comfortable for long dinner conversations, easy to move, and matches with a diverse range of dining tables.
        \item Able to stand pets' daily activities
    \end{itemize}
    \item Aesthetic Preferences:
    \begin{itemize}
        \item Prefers neutral tones with occasional bold accents.
        \item Likes clean lines and uncluttered spaces.
        \item Appreciates furniture that makes a statement but remains timeless.
    \end{itemize}
\end{itemize}

\subsubsection{Sketches \& Mood boards}
\begin{figure}[htbp]
    \centering
    \begin{subfigure}[t]{0.45\linewidth}
        \centering
        \includegraphics[width=\linewidth]{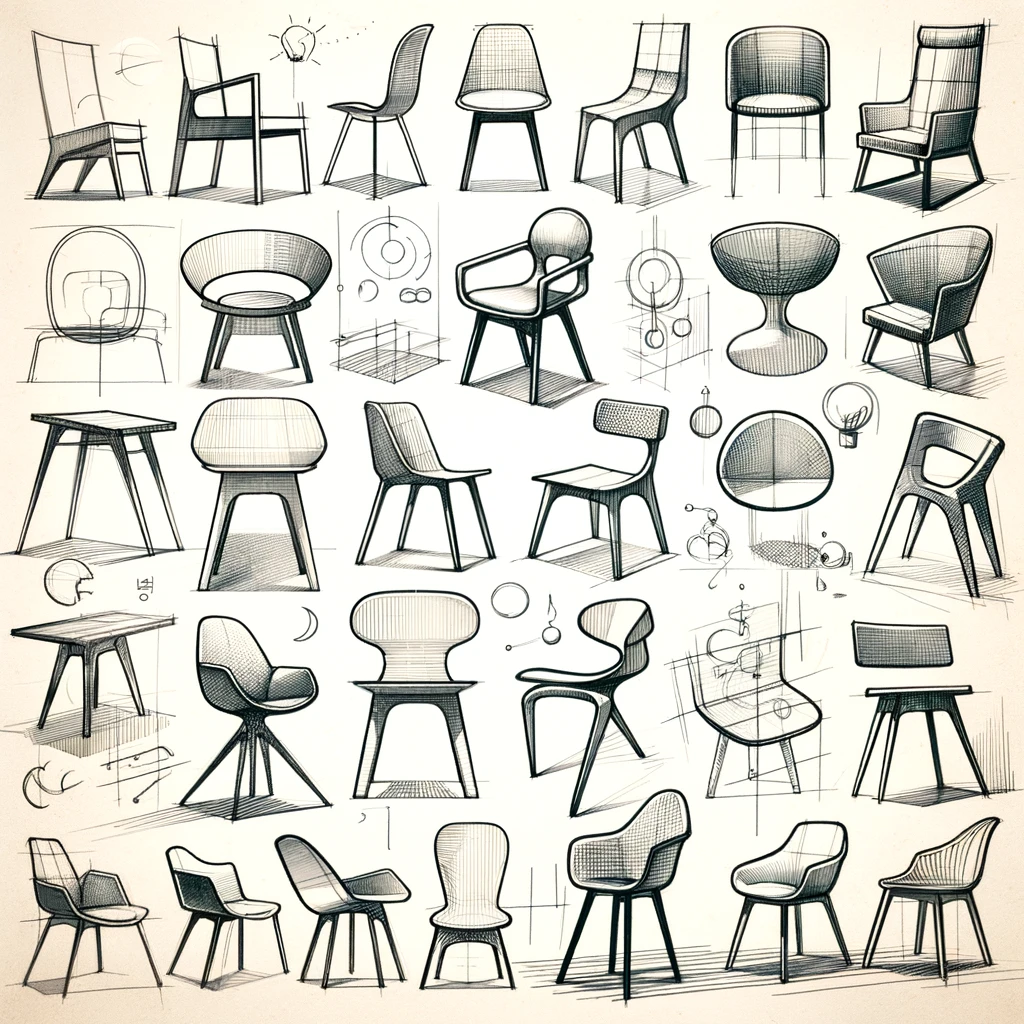}
        \caption{Design Document Sketches}
        \Description{An image with various sketches about chair designs}
        \label{fig:design_document_sketch}
    \end{subfigure}
    \hfill
    \begin{subfigure}[t]{0.45\linewidth}
        \centering
        \includegraphics[width=\linewidth]{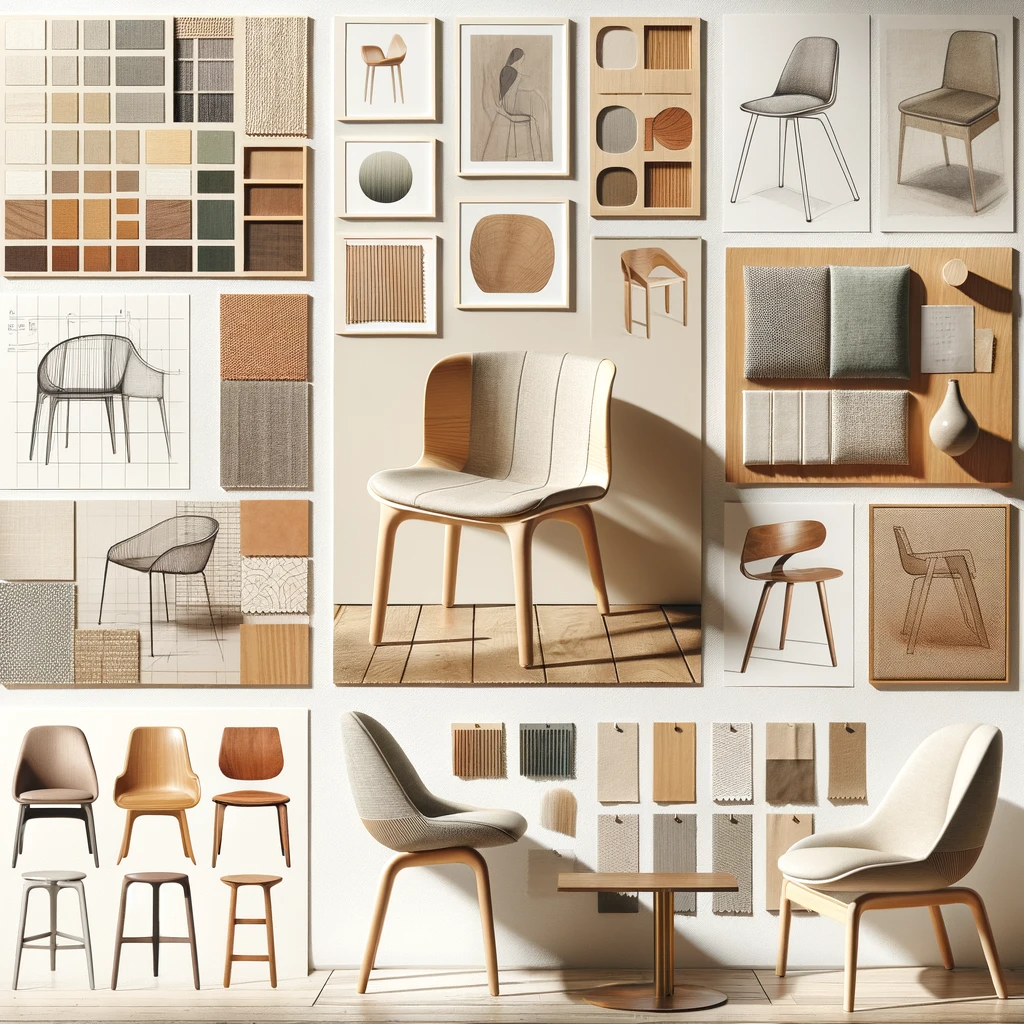}
        \caption{Mood Board 1}
        \Description{An image with cool design inspirations about chairs}
        \label{fig:design_document_moodboard1}
    \end{subfigure}
    
    \vspace{0.5cm} 
    \begin{subfigure}[t]{0.45\linewidth}
        \centering
        \includegraphics[width=\linewidth]{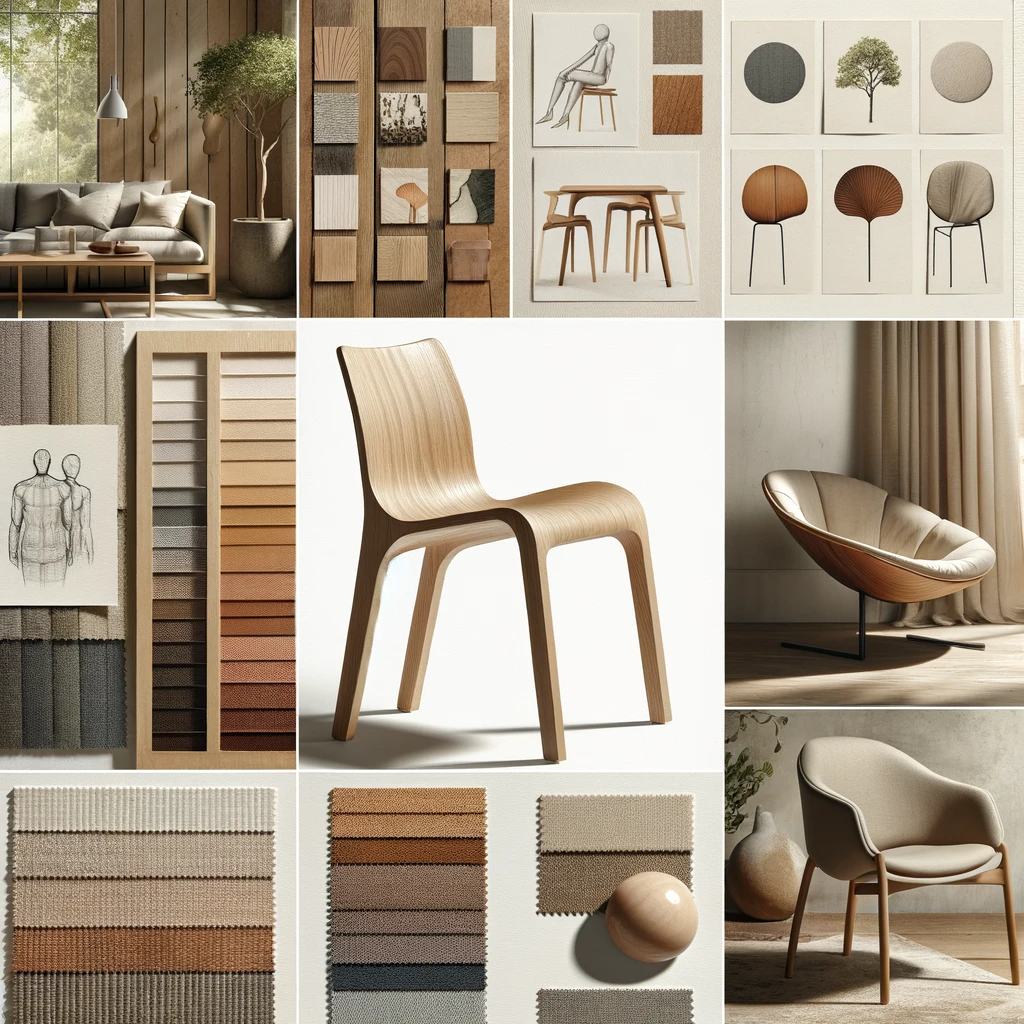}
        \caption{Mood Board 2}
        \Description{An image with cool design inspirations about chairs}
        \label{fig:design_document_moodboard2}
    \end{subfigure}
    \hfill
    \begin{subfigure}[t]{0.45\linewidth}
        \centering
        \includegraphics[width=\linewidth]{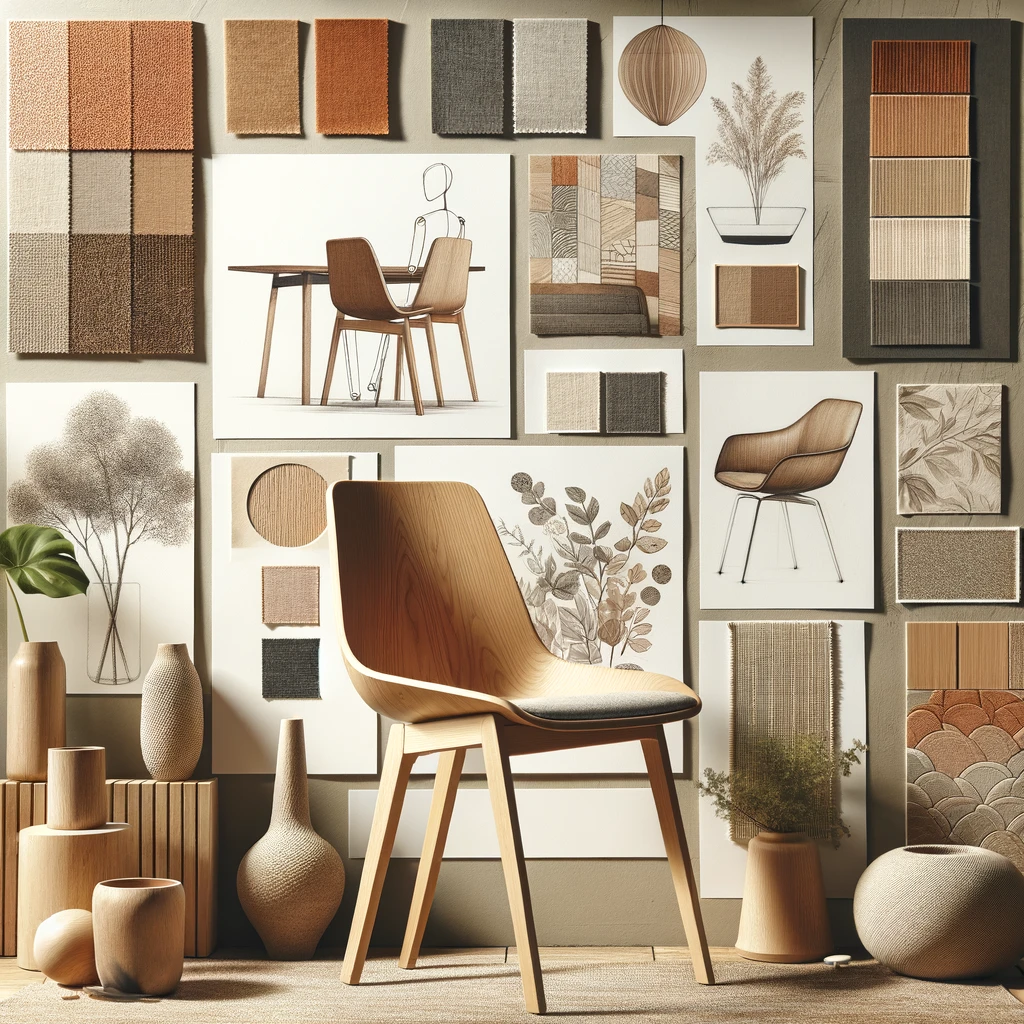}
        \caption{Mood Board 3}
        \Description{An image with cool design inspirations about chairs}
        \label{fig:design_document_moodboard3}
    \end{subfigure}

    \caption{Design Document: Sketches and Mood Boards}
    \label{fig:design_document_combined}
\end{figure}

Here we present the sketches and mood boards we showed to our participants in the design document (see \autoref{fig:design_document_combined}).

\section{Survey \& Interview Questions}
\label{appC:siq}
\subsection{Formative Study - Screening Survey}
\label{appC1:fs_ss}
\textbf{Design Interview Study}

Thank you for your interest in participating in our study! We are a group of researchers from the xxx at the xxx. We seek designers and creators with prior experience customizing products, solutions, or other client designs. Our goal is to understand authentic design workflows better.

Participants will receive \$xx for approximately 30 minutes of their time (\$yy/hour). The interview will be recorded for research purposes. If you are interested, please provide your background information and availability for the interview.

\subsubsection*{Participant Information}
\begin{enumerate}
    \item \textbf{Full Name:} \noindent\rule{3cm}{0.4pt}
    \item \textbf{Email Address:} \noindent\rule{3cm}{0.4pt}
\end{enumerate}

\subsubsection*{Design Experience}
\begin{enumerate}
    \item \textbf{What is your background in creating custom solutions in your domain?}  
    (e.g., Furniture, product design, graphics, architecture, etc.)  
    \begin{itemize}
        \item Please describe your experience, the types of products you design, and how long you have been in the field. 
    \end{itemize}
    \item \textbf{[Optional] Please share any links to your past projects or portfolio pieces that showcase your custom design work:} \noindent\rule{3cm}{0.4pt}
\end{enumerate}

\subsubsection*{Availability}
\begin{enumerate}
    \item \textbf{Preferred Interview Times} (check all that apply):
    \begin{itemize}
        \item \textbf{9 AM -- 12 PM:} Mon/ Tue/ Wed/ Thu/ Fri/ Sat/ Sun
        \item \textbf{12 -- 3 PM:} Mon/ Tue/ Wed/ Thu/ Fri/ Sat/ Sun
        \item \textbf{3 -- 6 PM:} Mon/ Tue/ Wed/ Thu/ Fri/ Sat/ Sun
        \item \textbf{Other}
    \end{itemize}

    \item \textbf{[Optional] Additional Comments or Questions} (e.g., time zone considerations)
\end{enumerate}

\subsection{Formative Study - Expert Interview Protocol}
\label{appC2:fs_eip}
\subsubsection*{Overview}
The following protocol outlines the structure and questions for conducting expert interviews as part of our research to better understand authentic design workflows.

\subsubsection*{Welcome Script}
\textbf{Month/Date @ Time - Interview with [NAME]}  

Hi [Interviewee Name],  

My name is xxx, and I am a researcher from the yyy Group at zzz Affiliation. Thank you so much for your interest and time in interviewing with us. We sincerely appreciate your help.  

To give you a brief overview, we are researching authentic design workflows to understand them better. We would love to ask you some questions to learn more about your background, work style, design process, and communication methods with clients.  

All of the information during this interview will be kept confidential, and with your consent, we might directly quote your insights. May we have your permission to record this interview for further analysis? The recording is intended to capture all the details you mention. Are you available to participate in this interview?  

\textbf{Permission to record interview:} [Yes/No]  
If yes: "Great! I will start the recording now."  

\subsubsection*{Interview Questions}
Can you tell me a little about your role and experience in the design field?

\textbf{Design Process}
\begin{enumerate}[label=\arabic*.]
    \setcounter{enumi}{1}
    \item Can you walk me through your typical design process, from ideation to completion?
        \begin{itemize}
            \item If possible, could you share details about a specific portfolio piece?
        \end{itemize}
    \item How do you handle client interactions and negotiations throughout the design process?
        \begin{itemize}
            \item E.g., trade-offs related to budget, aesthetics, and materials.
        \end{itemize}
\end{enumerate}

\textbf{Key Design Dimensions}
\begin{enumerate}[label=\arabic*.]
    \setcounter{enumi}{3}
    \item What are your key design considerations when designing a product?
        \begin{itemize}
            \item E.g., size, weight, depth, budget, materials, form, style.
        \end{itemize}
    \item Can you discuss a specific project where you faced challenges related to these design dimensions?
\end{enumerate}

\textbf{Negotiating with Clients}
\begin{enumerate}[label=\arabic*.]
    \setcounter{enumi}{5}
    \item How do you approach discussions with clients about design preferences?
        \begin{itemize}
            \item E.g., in-person, Zoom, Skype, Slack, Email.
        \end{itemize}
    \item Can you share an example of a trade-off situation and how you and the client resolved it?
        \begin{itemize}
            \item E.g., handling feedback and revisions.
        \end{itemize}
\end{enumerate}

\textbf{Client Constraints and Input Modalities}
\begin{enumerate}[label=\arabic*.]
    \setcounter{enumi}{7}
    \item What are some common constraints you encounter when working with clients?
    \item How do you manage and prioritize these constraints?
    \item How do clients typically provide input, and how do you incorporate this feedback into your designs?
        \begin{itemize}
            \item E.g., text descriptions, images, sketches.
        \end{itemize}
\end{enumerate}

\textbf{End}\\[0.5em]
Thank you so much for sharing your insights and experiences. Your input is incredibly valuable to our research. Could you also provide your preferred contact method for follow-up opportunities like user testing? Our research team will provide a [\$xx reward] for your participation as a token of appreciation. Do you have any questions or additional thoughts you would like to share before we conclude the interview?

\subsection{Summative Study - Screening Survey}
\label{appC3:ss_ss}
\subsubsection*{Participant Information}
\begin{enumerate}
    \item \textbf{Full Name:} \noindent\rule{3cm}{0.4pt}
    \item \textbf{Email Address:} \noindent\rule{3cm}{0.4pt}
    \item \textbf{Experiment Version:}
    \begin{itemize}
        \item In-person
        \item Remote
    \end{itemize}
\end{enumerate}

\subsubsection*{Demographic Information}
\begin{enumerate}
    \item \textbf{English Language Proficiency:}
    \begin{itemize}
        \item Native
        \item Advanced
        \item Intermediate
        \item Basic
        \item No proficiency
    \end{itemize}
    \item \textbf{Gender:}
    \begin{itemize}
        \item Female
        \item Male
        \item Non-binary/ Non-conforming
        \item Prefer not to respond
    \end{itemize}
    \item \textbf{Age:} \noindent\rule{3cm}{0.4pt}
\end{enumerate}

\subsubsection*{Design Experience}
\begin{enumerate}
    \item \textbf{Familiarity with Design:}
    \begin{itemize}
        \item Never
        \item Beginner (0–1 year)
        \item Intermediate (2–5 years)
        \item Advanced (>5 years)
    \end{itemize}
    \item \textbf{Design Fields (select all that apply):}
    \begin{itemize}
        \item Graphic Design
        \item Industrial Design
        \item UI/UX Design
        \item Architectural Design
        \item Fashion Design
        \item Other: \noindent\rule{3cm}{0.4pt}
    \end{itemize}
    \item \textbf{Describe Your Design Experience:} \noindent\rule{2cm}{0.4pt}
\end{enumerate}

\subsubsection*{Technical Experience}
\begin{enumerate}
    \item \textbf{Frequency of Using Large Language Models (e.g., ChatGPT, Gemini):}
    \begin{itemize}
        \item Never
        \item Tried a few times
        \item Monthly
        \item Weekly
        \item Daily
    \end{itemize}
    \item \textbf{Frequency of Using Image Generative Models (e.g., MidJourney, DALL-E):}
    \begin{itemize}
        \item Never
        \item Tried a few times
        \item Monthly
        \item Weekly
        \item Daily
    \end{itemize}
\end{enumerate}

\subsubsection*{Additional Information}
\begin{enumerate}
    \item \textbf{Do You Have Any Questions?} 
    \item \textbf{Participant ID:} Assigned ID (record this for future use).
\end{enumerate}

\subsection{Summative Study - Post-Experiment Survey}

\subsubsection*{Participant Details}
\begin{enumerate}
    \item \textbf{Participant ID:} \noindent\rule{3cm}{0.4pt}
    \item \textbf{Experiment Version:} \noindent\rule{3cm}{0.4pt}
    \begin{itemize}
        \item In-person
        \item Remote
    \end{itemize}
\end{enumerate}

\subsubsection*{Self-Evaluation}
\begin{enumerate}
    \item \textbf{What criteria did you use to select the final image for the client?}
    \item \textbf{Why does this image best fit your client's needs?} 
    \item \textbf{List the design dimensions and options you learned about the chair design space:} \noindent\rule{2cm}{0.4pt}
\end{enumerate}

\subsubsection*{Tool Feedback}
\begin{enumerate}[label=\arabic*.]
    \item \textbf{Image Satisfaction:}
    How satisfied were you with the generated images? (1=Very Unsatisfied, 7=Very Satisfied)  
    \emph{What factors influenced your rating?}
    
    \item \textbf{Image Alignment:}
    How well did the images match your expectations? (1=Very Misaligned, 7=Very Aligned)  
    \emph{Which elements aligned or deviated from your expectations?}
    
    \item \textbf{Ease of Converting Ideas to Prompts:}
    How easy was it to translate your ideas into prompts? (1=Very Difficult, 7=Very Easy)  
    \emph{What aspects made the process easier or more challenging?}
\end{enumerate}

\subsubsection*{Tool Usefulness}
\begin{enumerate}[label=\arabic*.]
    \item \textbf{Rate these statements} [1=Strongly Disagree, 7=Strongly Agree]
    \begin{itemize}
        \item The tool visualized my ideas.
        \item It helped explore design dimensions.
        \item It aided in initial prompt creation.
        \item It refined my design concept.
    \end{itemize}
    \item \textbf{Additional comments:} \noindent\rule{2cm}{0.4pt}
\end{enumerate}

\subsubsection*{Iterative Design Process}
\begin{enumerate}[label=\arabic*.]
    \item \textbf{Rate these statements} [1=Strongly Disagree, 7=Strongly Agree]
    \begin{itemize}
        \item My design improved over iterations.
        \item My prompts became more detailed.
    \end{itemize}
    \item \textbf{Additional comments:} \noindent\rule{2cm}{0.4pt}
\end{enumerate}

\subsubsection*{Future Directions}
\begin{enumerate}[label=\arabic*.]
    \item \textbf{Which aspects of the Design Assistant were most intuitive/useful and which were confusing/difficult?}
    \item \textbf{Suggestions for improvement:} \noindent\rule{2cm}{0.4pt}
\end{enumerate}

\section{Additional Analysis}
\begin{figure*}[htbp]
    \centering
    \includegraphics[width=\textwidth]{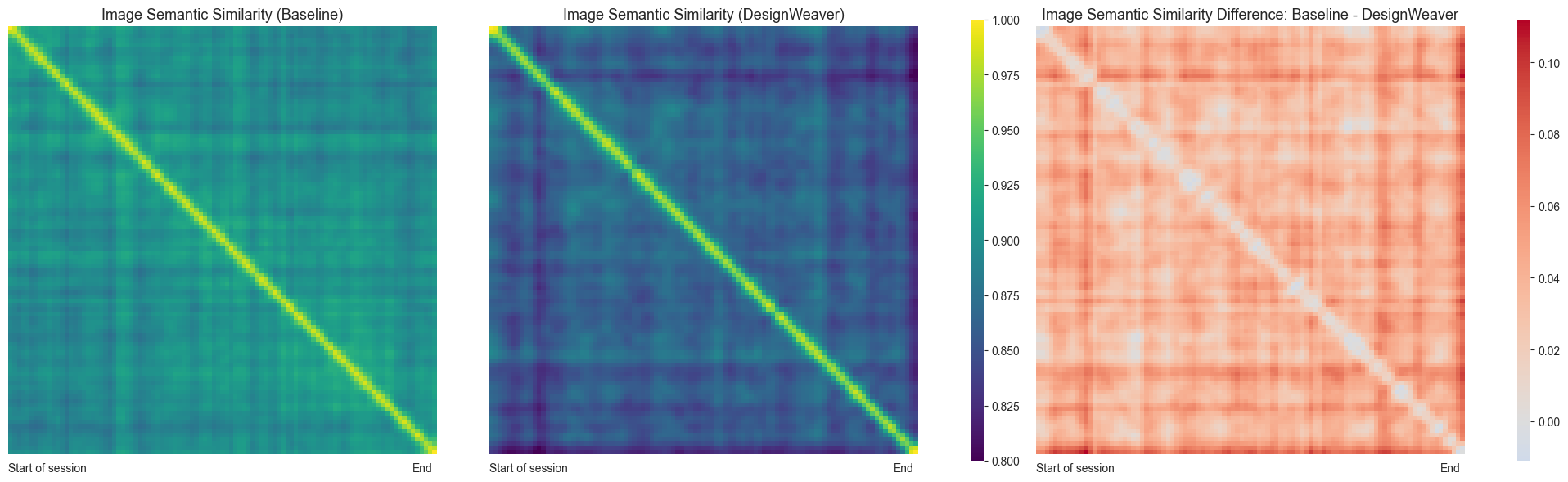}
    \caption{Image semantic similarity heatmaps for Baseline (left),  DesignWeaver (middle), and their differences (right). Baseline participants' generated images are more semantically similar.}
    \label{fig:finding_image_side_by_side_heatmaps}
    \Description{This figure includes three heatmaps. The first (left) represents the image-to-image correlations for the Baseline, illustrating how image outputs correlate across different iterations. The second (middle) shows the same for DesignWeaver, highlighting how DesignWeaver affects the image output process. The third (right) heatmap shows the correlations between Baseline and DesignWeaver, where red indicates areas of increased correlation and blue indicates areas of decreased correlation in the DesignWeaver group.}
\end{figure*}

\begin{figure*}[htbp]
    \centering
    \includegraphics[width=\textwidth]{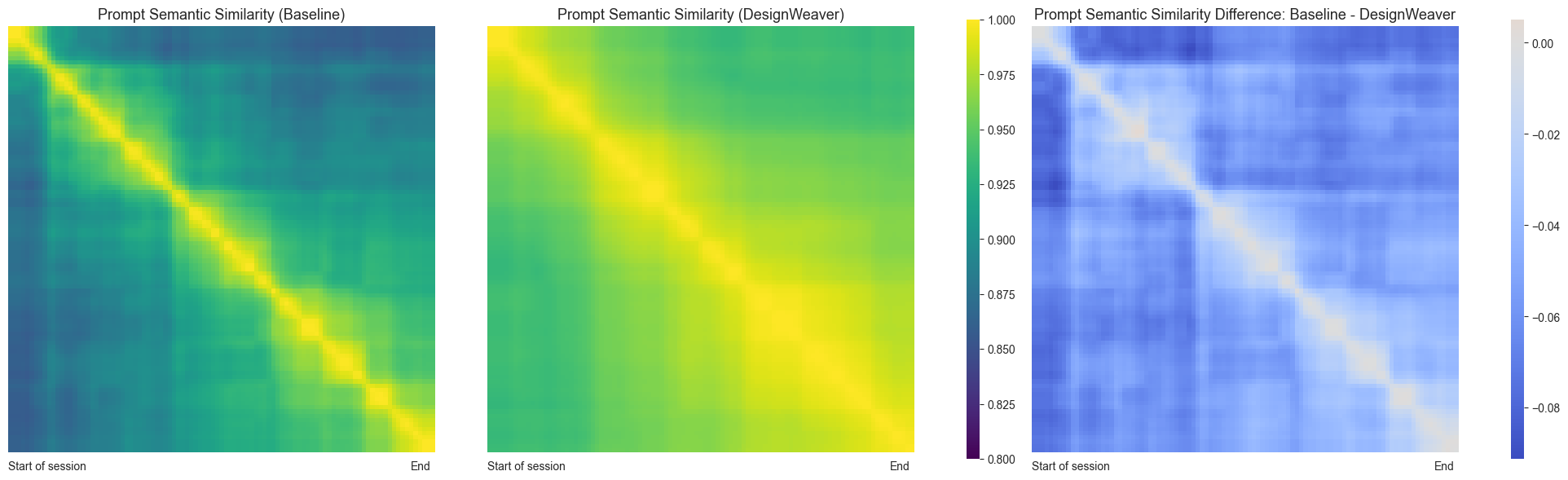}
    \caption{Prompt similarity heatmaps for Baseline (left),  DesignWeaver (middle), and their differences (right). DesignWeaver participants' prompts are more semantically similar.}
    \label{fig:finding_prompt_side_by_side_heatmaps}
    \Description{This figure includes three heatmaps. The first (left) represents the prompt-to-prompt Similarity for Baseline, showing how similar the prompts are across different iterations. The second (middle) represents the prompt Similarity for DesignWeaver, highlighting how the DesignWeaver process influences the Similarity between prompts over time. The third (right) shows the difference in prompt Similarity between Baseline and DesignWeaver, where blue areas indicate a decrease in Similarity and red areas indicate an increase in Similarity within the DesignWeaver group compared to the Baseline.}
\end{figure*}

\begin{figure*}[htbp]
    \centering
    \includegraphics[width=\textwidth]{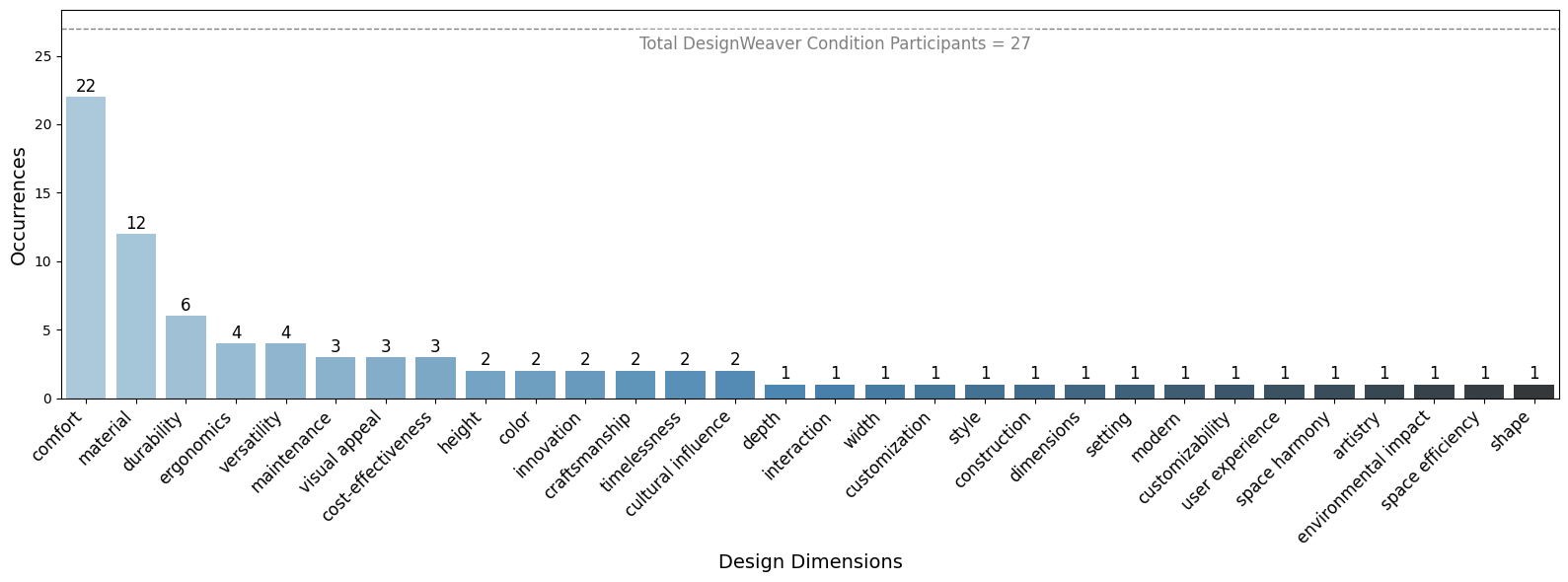}
    \caption{Frequency of design dimensions generated by DesignWeaver participants.}
    \label{fig:design_dimensions_frequency}
    \Description{
        The bar graph displays the count of each unique design dimension mentioned across participants in the DesignWeaver condition. The x-axis lists design dimensions (e.g., comfort, material, Durability), and the y-axis shows the count of occurrences for each dimension. Bars are labeled with the exact count of mentions. A dashed horizontal line at 27 represents the total number of participants in this condition, with an annotation reading "Total DesignWeaver Condition Users." The most frequent dimensions are comfort (22 mentions), material (12 mentions), and Durability (6 mentions), with others mentioned less frequently.
    }
    \label{fig:user_study_bar_chart}
\end{figure*}

\subsection{Prompt Comparison}
\label{appD:pc}

Here are randomly sampled example prompts from similar stages of the design process from \toolname{} and Baseline conditions (see \autoref{table:prompt_comparison}). 

\begin{table}[htbp]
\centering
\begin{tabular}{|p{0.32\linewidth}|p{0.66\linewidth}|} 
\hline
\textbf{Baseline \newline (P18 10th Iteration)} & \textbf{\toolname{} \newline (P7 7th Iteration)} \\ \hline
A dining chair that is beige in color, made out of wood and leather, with a unique leg design, a wide sitting space, an ergonomic design, a curved and tall back, and a minimalistic aesthetic. It is described as comfortable. & 
The design features a neutral vibe, bold accents, and contemporary elements. It includes unique, eye-catching details and emphasizes eco-friendly and durable elements such as sustainable materials, renewable resources, and energy-efficient production. Ergonomics, lightweight construction, sturdy build, and scratch-resistant surfaces further enhance its functionality. Wood adds natural warmth, and the curved back with dynamic posture support improves comfort. Playful geometry adds to the modern aesthetic. \\ \hline
\end{tabular}
\caption{Prompts generated in the \toolname{} condition are more developed in fewer iterations.}
\label{table:prompt_comparison}
\end{table}

\subsection{Semantic Difference}
\label{appD:sd}
\autoref{fig:finding_image_side_by_side_heatmaps} and \autoref{fig:finding_prompt_side_by_side_heatmaps} shows more fine-grained across iteration comparison of the semantic difference using CLIP. The figures indicate that \toolname{} has more semantically diverse visual outputs but more semantically similar prompts. We believe the more semantically similar prompts in \toolname{} condition could retrieve more semantically diverse images in the latent embedding space in the T2I model we used because they are more detailed and nuanced. One interesting pattern in \autoref{fig:finding_prompt_side_by_side_heatmaps} is that prompts are becoming more semantically similar as \toolname{} participants accumulate more tags in one prompt as time progresses.

\subsection{Design Dimensions}
\label{appD:dd}

\autoref{tab:designweaver_generated_dimensions} shows all the design dimensions 27 \toolname{} participants generated, which are 38 in total. After merging some with essentially the same meaning, we obtained the following \autoref{fig:design_dimensions_frequency}, which shows the frequency of newly generated design dimensions (other than the three dimensions we pre-populated to help with the cold start problem). These dimensions are all reasonable and align with the design document requirements.

\begin{table}[htbp]
    \centering
    \begin{tabular}{|p{0.95\linewidth}|} 
    \hline
    artistry, color, color palette, comfort, construction, cost-effectiveness, craftsmanship, cultural influence, customizability, customization, depth, dimensions, Durability, environmental impact, ergonomic design, ergonomics, height, innovation, interaction, maintenance, maintenance ease, material, material durability, material innovation, material quality, materiality, materials, modern, setting, shape, space efficiency, space harmony, style, timelessness, user experience, versatility, visual appeal, width. \\
    \hline
    \end{tabular}
    \Description{This table lists 38 distinct design dimensions alphabetically, including factors such as artistry, ergonomics, and material qualities.}
    \caption{List of 38 DesignWeaver Participant Generated Design Dimensions in Alphabetical Order}
    \label{tab:designweaver_generated_dimensions}
\end{table}

\end{document}